\documentclass{article}

\usepackage{arxiv}

\usepackage[utf8]{inputenc} 
\usepackage[T1]{fontenc}    
\usepackage{hyperref}       
\usepackage{url}            
\usepackage{booktabs}       
\usepackage{amsfonts}       
\usepackage{nicefrac}       
\usepackage{microtype}      
\usepackage{cleveref}       
\usepackage{lipsum}         
\usepackage{graphicx}
\usepackage{natbib}
\usepackage{doi}

\title{A seasonal climatology of the upper ocean pycnocline}

\author{Guillaume Sérazin \\
	    Ifremer, Univ. Brest, CNRS, IRD, \\
	    Laboratoire d’Océanographie Physique et Spatiale (LOPS)\\
        IUEM, 29280, Plouzané, France \\
	    \And
	    Anne Marie Tréguier \\
	    Ifremer, Univ. Brest, CNRS, IRD, \\
	    Laboratoire d’Océanographie Physique et Spatiale (LOPS)\\
        IUEM, 29280, Plouzané, France \\
	    \And
	    Clément de Boyer Montégut \\
	    Ifremer, Univ. Brest, CNRS, IRD, \\
	    Laboratoire d’Océanographie Physique et Spatiale (LOPS)\\
        IUEM, 29280, Plouzané, France \\
}

\renewcommand{\shorttitle}{Upper ocean pycnocline}

\hypersetup{
pdftitle={A template for the arxiv style},
pdfsubject={q-bio.NC, q-bio.QM},
pdfauthor={Guillaume Sérazin},
pdfkeywords={First keyword, Second keyword, More},
}

\graphicspath{{./Figures/}}
\newcommand{\dge}{$^\circ$E}
\newcommand{\dgw}{$^\circ$W}
\newcommand{\dgs}{$^\circ$S}
\newcommand{\dgn}{$^\circ$N}

\begin{document}
\maketitle

\begin{abstract}
Climatologies of the mixed layer depth (MLD) have been provided using several definitions based on temperature/density thresholds or hybrid approaches. The upper ocean pycnocline (UOP) that sits below the mixed layer base remains poorly characterised, though this transition layer is an ubiquitous feature of the ocean surface layer. Available hydrographic profiles provide near-global coverage of the world's ocean and are used to build a seasonal climatology of UOP properties -- intensity, depth, thickness -- to characterise the spatial and seasonal variations of upper ocean stratification. The largest stratification values $\mathcal{O}(10^{-3}\,s^{-2})$ are found in the intertropical band, where seasonal variations of the UOP are also very small. The deepest ($>$ 200 m) and least stratified  $\mathcal{O}(10^{-6}\,s^{-2})$ UOPs are found in winter along the Antarctic Circumpolar Current (ACC) and at high latitudes of the North Atlantic. The UOP thickness has a median value of 23 m with limited seasonal and spatial variations; only a few regions have UOP thicknesses exceeding 35 m. The UOP properties allow the characterisation of the upper ocean restratification that generally occurs in early spring and is generally associated with large variability. Depending on the region, this restratification may happen gradually as around the Rockall plateau or abruptly as in the Kuroshio Extension. The UOP is also likely to merge intermittently with the permanent pycnocline in winter. The upper limit of the UOP is eventually consistent with MLD estimates, except in a few notable regions such as in the Pacific Warm Pool where barrier layers are important, and during wintertime at high latitudes of the North Pacific.

\end{abstract}


\section{Introduction}

The upper ocean vertical structure can be decomposed into several layers based on density and stratification \citep{sprintall_upper_2009}. Near the surface, lies the ocean surface boundary layer (OSBL) that directly experiences the effect of air-sea exchanges. Buoyancy fluxes - i.e., heat fluxes, water evaporation and precipitation - as well as winds drive strong levels of turbulence that are associated with intense vertical mixing. Instabilities at the origin of such turbulent vertical motions include: convective instabilities due to destabilising buoyancy fluxes, wind-driven shear instabilities, breaking surface waves, Langmuir flows resulting from the interaction between the Stoke drift of surface waves and surface currents.

The action of this vertical mixing creates a zone of vertically homogeneous water called the mixed layer, whose depth can be estimated using a threshold in density or temperature to capture the rapid change in water properties that occurs at the base of this layer \citep[e.g.][]{de_boyer_montegut_mixed_2004, holte_argo_2017}. The mixed layer is distinguished from the mixing layer, which corresponds to the zone where mixing is currently active \citep{brainerd_surface_1995, johnston_observations_2009}. Although stratification within the mixed layer is weak, it can allow submesoscale baroclinic instability to trigger in the presence of horizontal density gradients \citep[e.g.,][]{boccaletti_mixed_2007, fox-kemper_parameterization_2008, callies_role_2016}. This so-called mixed layer instabilities play a key role in restratifying the upper ocean \citep{boccaletti_mixed_2007}.

Part of the OSBL is the so-called transition layer that connects the stratified, weakly turbulent interior, with the surface well-mixed layer. Mixing rates vary from high values in the mixed layer to extremely low values in the interior \citep{ferrari_eddy-mixed_2004}. This transition layer is also characterised by local maxima in stratification and shear \citep{johnston_observations_2009, sun_scaling_2013, kaminski_high-resolution_2021}. Using several definition of the transition layer, \cite{johnston_observations_2009} have shown that this layer can be well captured using the upper maximum of vertical stratification profiles along with other methods yielding a transition layer thickness from 8 to 24 m. Using a Lagrangian float equipped with a CTD, temperature chains and an ADCP, \cite{kaminski_high-resolution_2021} followed the evolution of the stratification, shear and turbulence of the transition layer during autumn in the north east area of the North Pacific. They found a thickness of 10-20 m consistent with \cite{johnston_observations_2009}. These previous estimates were, however, limited to a few regions and did not cover a full annual cycle, so that one cannot ascertain that the transition layer thickness is similar in other places of the global ocean and at any time of the year.  

Assessing the transition layer properties -- intensity, depth, thickness-- at the global scale is only possible using stratification inferred from historical hydrographic observations as autonomous floats provide a good coverage of the global ocean. Measuring turbulent dissipation and vertical shear is, however, only available in a few spots and requires dedicated field studies. Based on these hydrographic data, \cite{helber_temperature_2012} captured the transition layer using the absolute maximum in stratification and they computed the transition layer thickness using an estimate of the MLD as the upper boundary and the stratification half-maximum, taken over the whole hydrographic profile, as a lower boundary. Contrary to \cite{johnston_observations_2009}, they found very thick transition layer reaching more than 500 m thickness at midlatitudes. The definitions of the transition layer and its thickness used by \cite{helber_temperature_2012} are debatable, and we will suggest another definitions in our study.

Another local stratification maximum is usually found below the transition layer in the subtropical gyres and corresponds to the permanent pycnocline, whose seasonal variations are small. A local minimun in stratification is generally comprised between the upper and permanent pycnocline and characterises the so-called mode waters \citep{speer_chapter_2013, tsubouchi_comparison_2016}. Subtropical mode waters and the permanent pycnocline have been extensively mapped using hydrographic observations \citep{feucher_subtropical_2019}, based on a simple diagnostic method that captures two consecutive minimum and maximum in stratification \citep{feucher_mean_2016}. 

The spatial and temporal characteristics of upper ocean stratification may modulate certain ocean processes whose dynamics is much faster than seasonal timescales such as submesoscale features, internal waves and vertical mixing. \cite{erickson_seasonality_2018} suggest that the base of the mixed layer may no longer be well defined during wintertime because of the lack of a strong upper ocean pycnocline in the northeast midlatitude Atlantic Ocean \cite[see also][]{erickson_vertical_2020}. In this region, large submesoscale vertical velocities and buoyancy fluxes were shown to penetrate deeper than the mixed layer base in winter and spring \citep{yu_annual_2019}. Moreover, internal waves may be amplified near the surface due to the presence of a strong seasonal pycnocline \citep{lahaye_sea_2019}, with different internal tide regimes occurring from season to season due to the seasonality of the  upper ocean stratification \citep{barbot_background_2021}. Deep convection may occur if the upper ocean stratification is weak enough \citep[e.g.,][]{marshall_open-ocean_1999}. These processes are associated with vertical transports of tracers and nutrients, so that upper ocean stratification may therefore modulate ocean heat storage and biogeochemical activity through the aforementioned rapid processes \citep[e.g.,][]{levy_physical_2013, uchida_vertical_2020, bourgeois_stratification_2022}.

Climate change could yield substantial changes in upper ocean stratification, that would impact on exchanges between the air-sea interface and the ocean interior. Recent studies have indeed shown significant trends in upper ocean stratification using historical observations \citep{yamaguchi_trend_2019, li_increasing_2020, sallee_summertime_2021}. The way that upper ocean stratification is evaluated however varies between authors: \cite{sallee_summertime_2021} computed the stratification below the MLD in a constant layer of 15 m thickness, along with a more classical measure that is the density difference between the surface and 200 m depth \citep[e.g.][]{capotondi_enhanced_2012, somavilla_warmer_2017, yamaguchi_trend_2019}. None of these measures guarantee that the stratification peak associated with the transition layer is fully captured and the consistency of these methods still needs to be evaluated.

Global characterisations of the OSBL have been limited to the description of the mixed layer and its depth, while the upper ocean pycnocline (UOP) that sits below the mixed layer base, sometimes referred to as the transition layer (see earlier in the introduction), remains poorly characterised. In this study, we are interested in estimating the UOP directly from hydrographic profiles, with the goal of building a mapped climatology of the UOP. With the number of past in situ oceanic observations, it is now possible to obtain an accurate description of the upper ocean vertical structure to address the following questions at the global scale:

\begin{itemize}
  \item How seasonally and spatially variable are the amplitude and depth of the UOP? 
  \item Can the UOP be considered as a layer with constant thickness?
  \item Are the UOP characteristics useful to discuss the timing and the variability of restratification periods?
\end{itemize}

The first question aims to complement the seasonal and geographical variability of the OSBL, usually discussed with estimates of the MLD. Moreover, it will complement the climatology of the permanent pycnocline and mode waters provided by \cite{feucher_subtropical_2019}, which only concerns subtropical gyres. The second question is motivated by the limited knowledge of this layer at the global scale, while some studies have made the assumption that the upper ocean pycnocline had a constant thickness \cite[e.g.][]{sallee_summertime_2021}. The last question is motivated by the capacity of submesoscale motions to penetrate beneath proxys of the mixed layer base when the stratification is weak. This question is also of interest because spring restratification is not a continuous process and may have an intermittent behaviour before the OSBL becomes fully restratified in summer.

The article is organised as follows. Section~\ref{sec:data_method} presents the hydrographic data and the method used to detect the upper ocean pycnocline. Section~\ref{sec:descriptive_approach} consists in a descriptive approach of the upper ocean stratification from a gridded temperature and salinity climatology. Section~\ref{sec:seasonal_climatology} describes the summer and winter climatologies of the UOP characteristics, their seasonal amplitude, and characterises restratification periods using the UOP. Section~\ref{sec:regional_analyses} focuses on a few regions to describe in more detail their stratification profiles, the associated UOP seasonal cycle along with the summer and winter variability, and the restratification in spring. Section~\ref{sec:discussion} discusses the UOP position relative to MLD estimates and to the permanent pycnocline. The UOP is also compared to classical measures of upper ocean stratification, and dynamical processes relevant to explain the UOP thickness are discussed. Section~\ref{sec:conclusion} draws the conclusions of our study.

\section{Data and methods}\label{sec:data_method}

We use the ISAS20\_ARGO dataset containing Argo and Deep-Argo temperature and salinity profiles on the period 2002-2020. Vertical profiles are available on 187 standard depth levels between 0 and 5500 m depth, with higher vertical resolution close to the surface (2-5 m) and coarser resolution in the deep ocean (100 m), resulting from the interpolation performed with the ISAS procedure \citep{gaillard_situbased_2016}. We select data in the depth range 0-2000 m as only recent floats provides measurement below 2000 m. In the first part of this study (section~\ref{sec:descriptive_approach}), we will use the gridded seasonal climatology of temperature and salinity fields, which is the result of an optimal interpolation applied on the profile set \citep[see for a full description of the interpolation method][]{gaillard_situbased_2016}. In the rest of the study, we will operate directly on individual profiles to compute the properties of the UOP.

\begin{figure}[h!]
  \begin{center}
    \includegraphics[width=11cm]{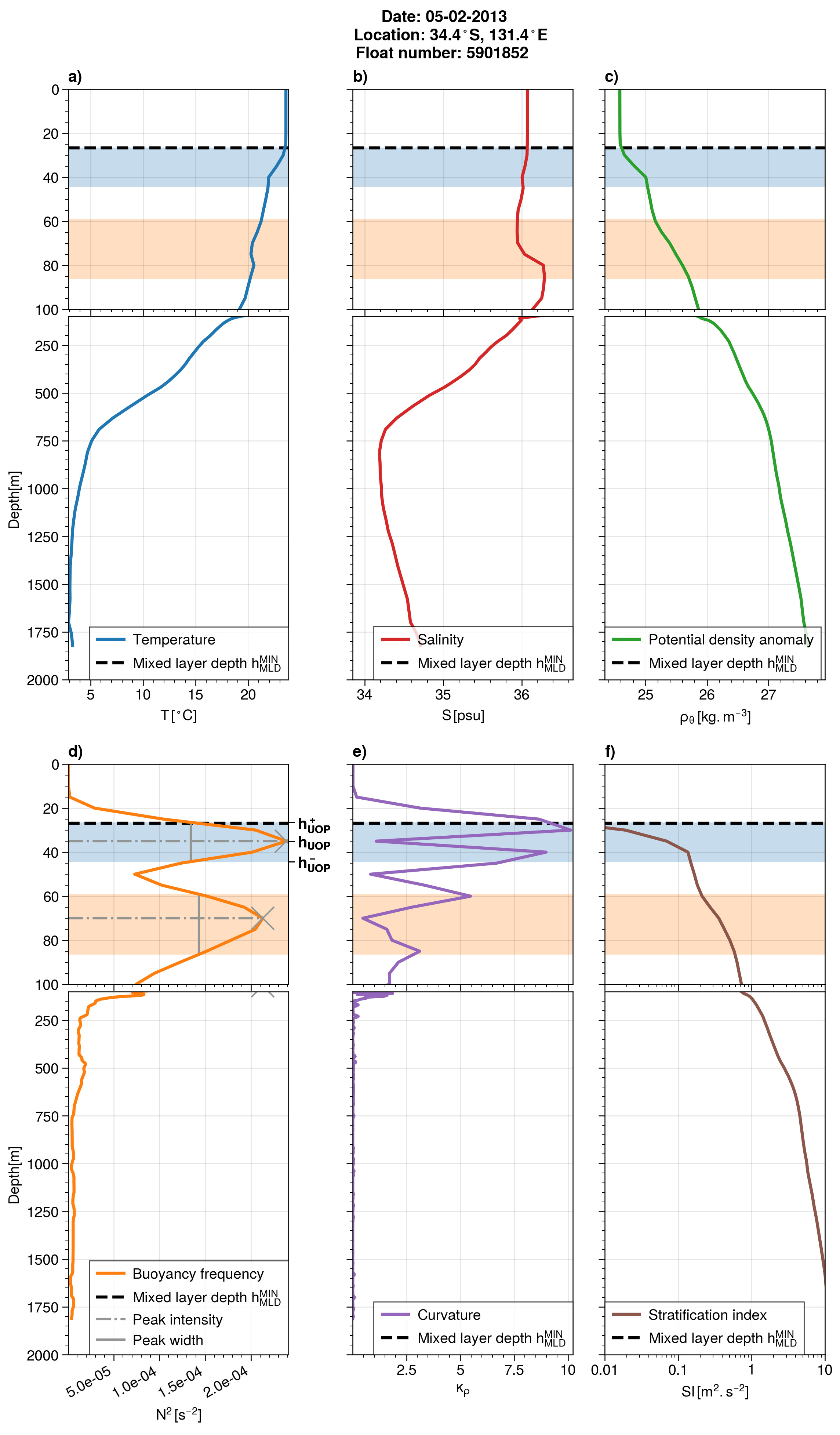}
  \end{center}
  \caption{Example of an hydrographic profile in the South Atlantic ocean (34 \dgs , 58.6 \dgw) recorded on 5 February 2013 with a) in situ temperature, b) practical salinity, c) potential density anomaly referenced to 0 dbar. Quantities derived from the density profile are: d) the buoyancy frequency squared e) the density curvature an f) the stratification index. Stratification peaks detected by the peak algorithm and their properties are shown in d) and the area corresponding to the peaks is represented by color shadings. The first peak is the UOP; the upper limit $h_{UOP}^{+}$ and the lower limit $h_{UOP}^{-}$ correspond to the edges of the shaded blue area. The minimum MLD of three variables $h_{MLD}^{MIN}=min(h^{\Delta \rho_\theta}_{MLD}, h^{\Delta T}_{MLD}, h^{\Delta \rho_\theta \propto \Delta T}_{MLD})$, described in section~\ref{sec:data_method}, is plotted with dashed lines.}
  \label{fig:profile_example}
\end{figure}

Profiles taken from ISAS20\_ARGO are linearly interpolated onto regular vertical levels with 5 m bins. A profile example evaluated on these vertical levels is shown for in situ temperature $T$ and practical salinity $S_P$ (Figure~\ref{fig:profile_example}a,b). The potential density referenced to 0 dbar $\rho_\theta$ is computed using the Gibbs seawater function from TEOS-10 \citep{mcdougall_getting_2011} as shown in the profile example of Figure~\ref{fig:profile_example}c. In the example, the temperature and salinity profiles show a well-homogenised mixed layer above 30 m depth (Figure~\ref{fig:profile_example}a,b) that is reflected by a well-defined mixed layer on potential density (Figure~\ref{fig:profile_example}c). 

We then compute three quantities derived from the surface-referenced potential density profile. First, the Brunst-Vaisala frequency squared $N^2$ is evaluated as
\begin{equation}\label{eq:n2}
 N^2(z) = \frac{g}{\rho_0}\frac{d\rho_\theta}{dz}
\end{equation}
where $z$ is the depth (positive downward), $g$ is the gravitational acceleration taken as 9.81 $m^2.s^{-1}$, and $\rho_0$ is the reference density taken as 1025 $kg.m^{-3}$. We use a 5-point Savitzky-Golay filter \citep{savitzky_smoothing_1964} to compute smoothed versions of the vertical derivatives of density, while preserving peaks. Results given by the Savitzky-Golay filter were compared to those using a 5-point running mean operator followed by a usual derivation: as expected the Savitzky-Golay filter better preserves the peaks in the signal. The stratification $N^2$ of the profile example clearly shows two stratification peaks (Figure~\ref{fig:profile_example}d), with the first one corresponding to the density jump at the base of the mixed layer, i.e. the UOP.

Stratification peaks are detected in smoothed $N^2$ profiles using the Python algorithm \textit{scipy.signal.find\_peaks} with the following parameters. Peaks are detected when their amplitude exceed $2 \sigma$, with $\sigma$ being the vertical standard deviation of the stratification profile $N^2$. The minimum prominence of peaks, that is the elevation between the peak and the surrounding troughs must be larger than $\sigma$. In practice, we limit the evaluation of the peak prominence within a window of 250 m centered around the peaks instead of using the full profile. This allows more efficient computation. The minimum distance between peaks is set to 25 m, which corresponds to a rough guess for the transition layer / UOP thickness \citep[e.g.,][]{johnston_observations_2009}. Below this distance, the two peaks are considered to be only one wide peak with the largest peak setting the amplitude. The minimum width of peaks is not constrained, so that we allow the detection of sharp peaks. From the finding of stratification peaks, the UOP is defined as being the closest peak to the surface. The UOP intensity is defined as the peak amplitude, the UOP depth as the peak depth, and the UOP thickness is computed at half height as illustrated in Figure~\ref{fig:profile_example}d. The upper and lower UOP limits at half height are termed $h_{UOP}^{+}$ and $h_{UOP}^{-}$, respectively. Note that our definition is strongly different from the transition layer defined by \cite{helber_temperature_2012}, who based their analysis on the MLD and on the deepest $N^2$ half-maximum over the full profile, hence estimating most of the time a much
thicker value than we do for this layer. Here, our definition is independent from the MLD definition and we focus on the first $N^2$ maximum from the surface.

Secondly, we compute the density curvature as
\begin{equation}\label{eq:kappa}
\kappa_{\rho_\theta}(z) = \frac{\left|  \frac{d^2 \hat{\rho}}{d\hat{z}^2} \right|}{\left[ 1 + \left(\frac{d\hat{\rho}}{d\hat{z}}\right)^2 \right]^{3/2}}, \mbox{ with } \hat{\rho}=\frac{\rho_\theta}{\rho_0} \mbox{ and } \hat{z}=\frac{z}{H}.
\end{equation}
Here, we use $H$ equals to the maximum depth studied, i.e., 2000 m. This quantity shows local maxima when the density profile has substantial bends, i.e. rapid change in slope, as illustrated by the profile example (Figure~\ref{fig:profile_example}e). In this example, curvature maxima are also close to peak edges estimated at mid height, with the first curvature maximum corresponding to the upper limit of the UOP ($h_{UOP}^{+}$). We performed this comparison for all the profiles and found that the maximum curvature above the UOP depth is very close to the upper limit of the UOP: a test performed on all the profiles showed that for less than 5\% of the profiles, the upper limit of the UOP was deeper than 9 m compared to the maximum curvature, for less than 5\% the upper limit of the UOP was shallower than 7 m compared to the maximum curvature ; the median difference between both limits was 1 m. Thus, the upper limit of the UOP is consistent with the geometric consideration of the density curvature. Note that \citet{lorbacher_ocean_2006} attempted to define a criteria to evaluate the MLD based on the  curvature of the temperature profile.

Thirdly, we compute a stratifcation index down to a given depth $h$:
\begin{equation}\label{eq:si_n2}	
  SI(h) = \int_0^{h} N^2(z)zdz.
\end{equation}
This index, sometimes called the columnar buoyancy, has been popular for studying deep convection in Mediterranean studies \cite[e.g.,][]{herrmann_modeling_2008, bosse_scales_2016, somot_characterizing_2018}. This stratification index corresponds to the total amount of buoyancy loss required to destratify the water column up to a depth $h$, implying that vertical mixing could extend up to this depth and homogenise water masses. Using equation (9.2.5) in \cite{turner_buoyancy_1973} \cite[see also][]{lascaratos_high-resolution_1998, herrmann_modeling_2008}, one may show that temporal variations in the stratification index $SI$ are linked to the intensity of surface buoyancy fluxes $B_{surf}$ and lateral buoyancy fluxes $B_{lat}$ by the following equation:
\begin{equation}\label{eq:dsi_dt}	
  \frac{\partial SI(h)}{\partial t} = N^2(h)h \frac{\partial h}{\partial t} = B_{surf} + B_{lat}.
\end{equation}

Equation \ref{eq:si_n2} can also be rewritten so that the stratification index $SI$ can be computed directly by integrating the density field as:  
\begin{equation}\label{eq:si_rho}	
  SI(h) = \frac{g}{\rho_0}\left[h\rho_\theta(h) - \int_0^h \rho_\theta(z)dz \right].  
\end{equation}
We use this form to compute $SI$ in our process as no vertical derivation is needed. In the profile example, the stratification index is obviously very close to zero within the mixed layer as the stratification is very weak. It starts to increase at the base of the mixed layer with the presence of the stratification peak (Figure~\ref{fig:profile_example}f).

We define the upper ocean stratification index $UOSI$, which is the previous stratification index estimated at the lower boundary of the UOP such that $UOSI = SI(h_{UOP}^{-})$. The UOSI is thus the combination of the UOP intensity, depth and thickness. For instance, a deep strongly-stratified UOP will have a large UOSI, meaning that large amount of buoyancy loss are required to erode the upper ocean stratification. On the contrary, a shallow weakly-stratified UOP will have a small UOSI, meaning that the upper ocean stratification will be easily eroded by buoyancy loss. Large variations in the UOSI would also denote the importance of buoyancy fluxes as shown by equation \ref{eq:dsi_dt}. We aim to use this quantity as a proxy for finding restratification periods when the $UOSI$ is small and shallow weakly-stratified layers are formed. Note that the UOSI may also be though as the upper ocean memory of past surface and lateral buoyancy forcing. 

The MLD is estimated using different threshold methods as in \cite{de_boyer_montegut_mixed_2004}. For all MLD variables, the reference state is taken at 10 m depth. $h^{\Delta \rho_\theta}_{MLD}$ is defined at the depth where the density becomes larger than a threshold of 0.03 $kg.m^{-3}$; $h^{\Delta T}_{MLD}$ uses a threshold of 0.2$^\circ$C on conservative temperature; $h^{\Delta \rho_\theta \propto \Delta T}_{MLD}$ uses a variable density threshold corresponding to a temperature threshold of 0.2$^\circ$C, therefore cancelling the salinity effects on density stratification and taking into account the local conditions (for instance, the 0.03 $kg.m^{-3}$ threshold value corresponds to a 0.2$^\circ$C threshold at a temperature of 8$^\circ$C and a salintiy of 35 PSU); $h^{MIN}_{MLD}$ is the minimum of all these MLD variables. Including the variable density threshold criteria with $h^{\Delta \rho_\theta \propto \Delta T}_{MLD}$ avoid to overestimate the MLD in polar regions \cite[e.g.,][]{courtois_mixed_2017, piron_argo_2016}, while including the temperature threshold criteria with $h^{\Delta T}_{MLD}$ allows the detection of compensated layers.

The relative position between the top of the UOP ($h_{UOP}^{+}$) and the MLD estimates will be discussed at the end of this paper. We also include in our analysis the climatology of the permanent ocean pycnocline (POP) produced by \cite{feucher_subtropical_2019} to discuss potential mergers between the UOP and the POP.

The UOP characteristics evaluated on each profile are gathered in 2$^\circ\times$2$^\circ$ bins for each month of the year, in which the median is taken to produce a gridded UOP climatology. The results are smoothed using a diffusive gaussian filter with a 500 km scale \citep{grooms_diffusion-based_2021, loose_gcm-filters_2022}. In this study, we mainly present the gridded UOP climatology for winter and summer, respectively defined as  January-February-March (JFM) and July-August-September (JAS) for the northern hemisphere. We add spring (March-April-May, MAM) and autumn (October-November-December, OND) for the UOSI. The extended monthly climatology is shown in supplementary materials. The corresponding dataset is made widely available, including the UOP gridded fields and the UOP characteristics evaluated on each profile.

\section{Phenomenology from a gridded climatology}\label{sec:descriptive_approach}

\begin{figure}[h!]
  \begin{center}
    \includegraphics[width=\textwidth]{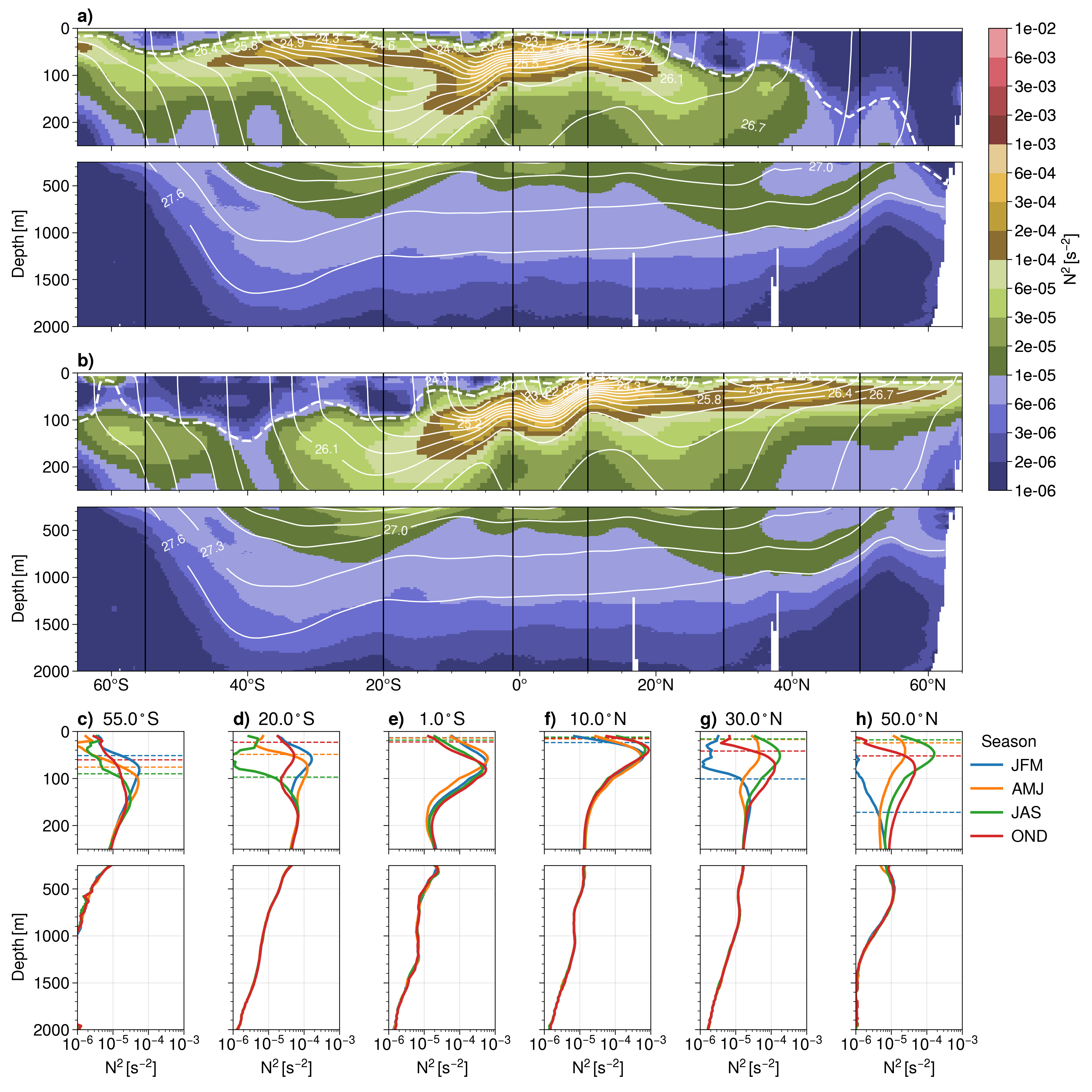}
  \end{center}
  \caption{Mean vertical stratification $N^2$ from the ocean surface down to 2000 m for a section in the Atlantic ocean at 25\dgw\ for a) JFM and b) JAS. The stratification is plotted with colour contours using a logarithmic scale, isopycnal surfaces are represented with labelled white lines and the mixed layer depth $h_{MLD}^{\Delta \rho_\theta}$ calculated with a density threshold of 0.03 $kg.m^{-3}$ is plotted with a thick dashed white line. Six profiles are plotted for the four seasons at different latitudes: 55\dgs, 20\dgs, 1\dgs, 10\dgn, 30\dgn, 50\dgn. The MLD $h_{MLD}^{\Delta \rho_\theta}$ for each season is plotted with horizontal dashed lines on the profiles.}
  \label{fig:n2_climatology_atlantic}
\end{figure}

\begin{figure}[h!]
  \begin{center}
	\includegraphics[width=\textwidth]{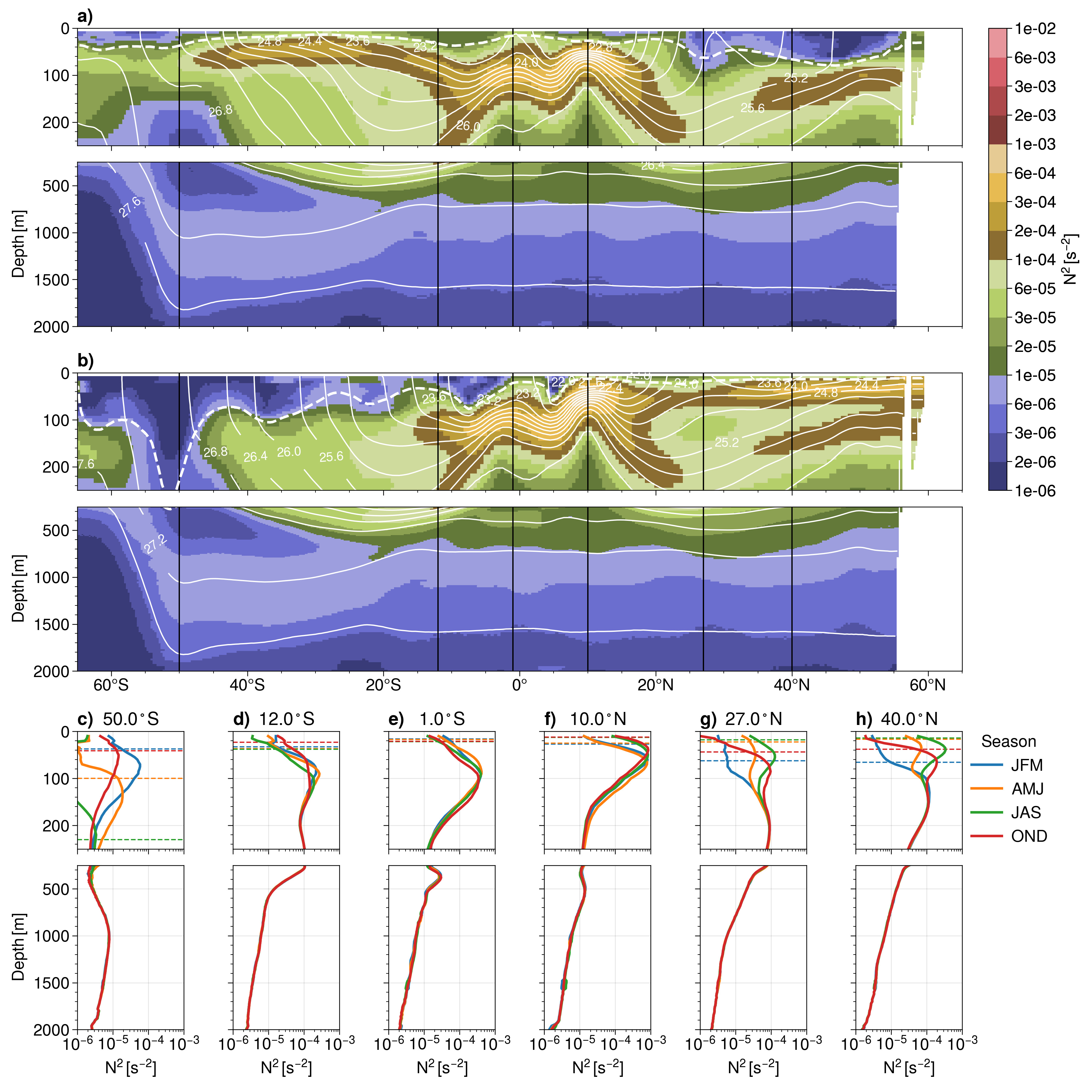}
  \end{center}
  \caption{Same as Figure~\ref{fig:n2_climatology_atlantic} but for a section in the Pacific ocean at 135\dgw\ and profiles at latitudes 50\dgs, 12\dgs, 1\dgs, 10\dgn, 27\dgn, 40\dgn.}
  \label{fig:n2_climatology_pacific}
\end{figure}

Using the gridded seasonal climatology of temperature and salinity, we first illustrate the meridional variations of upper ocean stratification with two meridional sections for the summer and winter seasons: one section in the Atlantic basin at 25\dgw\ (Figure~\ref{fig:n2_climatology_atlantic}a,b) and one section in the Pacific basin at 135\dgw\ (Figure~\ref{fig:n2_climatology_pacific}a,b). In the following, we will use three arbitrary levels of stratification to discuss the ocean vertical structure: largely stratified waters with $N^2 \geq 10^{-4}s^{-2}$, moderately stratified waters with $10^{-5}s^{-2}<N^2<10^{-4}s^{-2}$ and weakly stratified waters with $N^2 \leq 10^{-5}s^{-2}$. 

Large stratification values are found right under the weakly-stratified mixed layer in summer at mid-to-high latitudes, and in the intertropical band between 15\dgs\ and 20\dgn\ for all seasons (Figures~\ref{fig:n2_climatology_atlantic}~and~\ref{fig:n2_climatology_pacific}). These largely stratified waters are identified with a local maximum that corresponds to the UOP. In general, the stratification tends to decrease below this stratification maximum and reaches weak values (i.e., less than $10^{-5}s^{-2}$) below 750 m. In the oceanographic literature, the region of moderately stratified waters is sometimes referred to as the main pycnocline, while waters below this depth are usually termed deep waters. Here, we prefer to use the definition of the pycnocline as a local maximum in the stratification profile since it does correspond to significant variations in density with depth.

In the intertropical band, a largely stratified pycnocline, corresponding to the UOP, is present below the mixed layer, with its intensity and depth fluctuating with the season. Seasonal variations are, however, small and the pycnocline remains largely stratified throughout the year. Another pycnocline of moderate stratification may be found at around 300-400 m at 1\dgs\ (Figure~\ref{fig:n2_climatology_atlantic}e, Figure~\ref{fig:n2_climatology_pacific}e) and to a lesser extent at 10\dgn\ (Figure~\ref{fig:n2_climatology_pacific}f). 

At higher latitudes, one may find a double peak configuration with a seasonal pycnocline, corresponding to the UOP, whose depth and amplitude vary substantially with the season, and a permanent pycnocline whose properties remain unchanged throughout the year. Depending on the latitude, the peak corresponding to the seasonal pycnocline may deepen and weaken in intensity (Figure~\ref{fig:n2_climatology_pacific}d), or it may vanish by weakening and merging with the permanent pycnocline below to form a unique pycnocline (Figure~\ref{fig:n2_climatology_atlantic}d,h and Figure~\ref{fig:n2_climatology_pacific}c,g,h). When the upper pycnocline weakens or vanishes, the permanent pycnocline becomes the UOP and the part of the ocean interior that sits above may connect with the surface ocean. This configuration allows the formation of mode waters comprised between the permanent and the seasonal pycnocline, with a mode water corresponding to a stratification minimum between the two maxima \cite[e.g.,][]{feucher_mean_2016}. This configuration generally happens within subtropical gyres of the ocean basins \cite[e.g.,][]{feucher_subtropical_2019} as at 20\dgs\ in the Atlantic section (Figure~\ref{fig:n2_climatology_atlantic}d) and at 12\dgs\ and 27 \dgn\ in the Pacific section (Figure~\ref{fig:n2_climatology_pacific}d,g), but it may also occur closer to the pole as in the Pacific section at 50\dgs\ (Figure~\ref{fig:n2_climatology_pacific}c).

Another configuration is found with only one seasonal pycnocline present in the upper ocean, associated with a gradual decrease in stratification with depth, so that the upper ocean stratification peak becomes less pronounced in winter. In this case, mode waters are absent and the seasonal pycnocline corresponds to the UOP at all times of year. This configuration happens for instance at 55\dgs\ and 30\dgn\ in the Atlantic section with a substantial deepening and weakening of the stratification peak for the latter section (Figure~\ref{fig:n2_climatology_atlantic}c,g).

Overall, the largest pycnocline over the depth range 0-2000 m remains largely to moderately stratified within the subtropical gyres. On the contrary, the pycnoclines may become weakly stratified in winter as one gets closer to the pole so that the full water column is weakly stratified during this season over the depth range 0-2000 m. It may happen with a double pycnocline configuration with the upper pycnocline vanishing in winter and the permanent pycnocline being weakly stratified (Figure~\ref{fig:n2_climatology_atlantic}h, Figure~\ref{fig:n2_climatology_pacific}c).
 
The description of these two sections of the seasonal gridded stratification $N^2$ confirms that the UOP is large at the equator and has small seasonal variations, whereas it may vary substantially at higher latitudes and may become weakly stratified in winter. This phenomenological approach also highlights the variety of upper ocean stratification configurations with double peak or single peak structures, and large flavours in seasonal variations.

\section{Global climatology of UOP characteristics}\label{sec:seasonal_climatology}

The previous descriptive approach provides insights on the seasonal and spatial behaviour of upper ocean stratification, but the UOP and the intensity of the associated stratification barrier must be captured on raw profiles first before being averaged to produce an accurate climatology. This method is indeed not equivalent to computing the UOP directly from gridded temperature and salinity climatologies due to nonlinear effects within the equation of state of seawater and those of the peak detection method. In this section, results from the gridded climatology of UOP characteristics are shown following the UOP detection on each hydrographic profile as detailed in the methodological part of this study (section~\ref{sec:data_method}).

\subsection{Winter and summer climatology}

From the gridded climatology, we build the distribution at each latitude of the three main characteristics of the gridded UOP -- intensity, depth and thickness. This captures the main meridional variations of the UOP characteristics, which are shown for the winter and summer seasons in Figure~\ref{fig:meridional_variations}. Winter and summer maps are also shown in Figure~\ref{fig:uop_characteristics} to display the spatial variations of the three UOP characteristics.

\subsubsection{Intensity}

\begin{figure}[h!]
  \begin{center}
    \includegraphics[width=\textwidth]{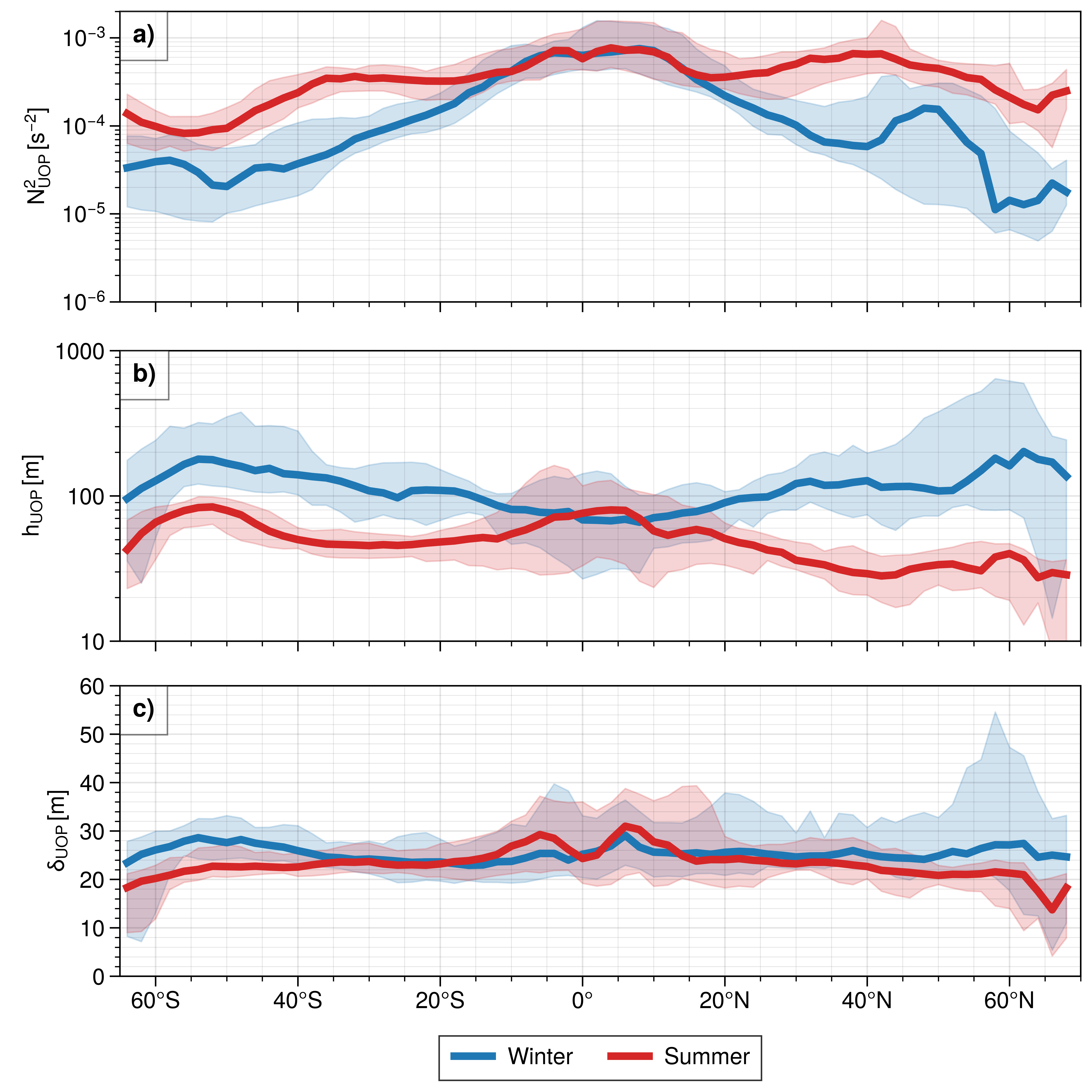}
  \end{center}
  \caption{Meridional distribution of median UOP characteristics for the winter season (blue curves) and for the summer season (red curves): a) intensity $N^2_{UOP}$, b) depth $h_{UOP}$ and c) thickness $\delta_{UOP}$. The shaded areas represent the amplitude of the distributions within each zonal band based on the 5 and 95 percentiles.}
  \label{fig:meridional_variations}
\end{figure}

The median summer UOP stratification is large and of the same order of magnitude between the equator and midlatitudes ($\sim 2-6.10^{-4}\,s^{-2}$), but becomes smaller south of 40\dgs\ in the Southern Ocean (Figure~\ref{fig:meridional_variations}a). The winter UOP intensity, however, has a notable decrease from the equator with large stratification values $\mathcal{O}(10^{-4}\,s^{-2})$ to the pole with moderate stratification values $\mathcal{O}(10^{-5}\,s^{-2})$. This leads to a seasonal amplitude increasing with latitude outside the intertropical band (20\dgs\ -- 20\dgn ), where both summer and winter stratification are comparable.

\begin{figure}[h!]
  \begin{center}
    \includegraphics[width=\textwidth]{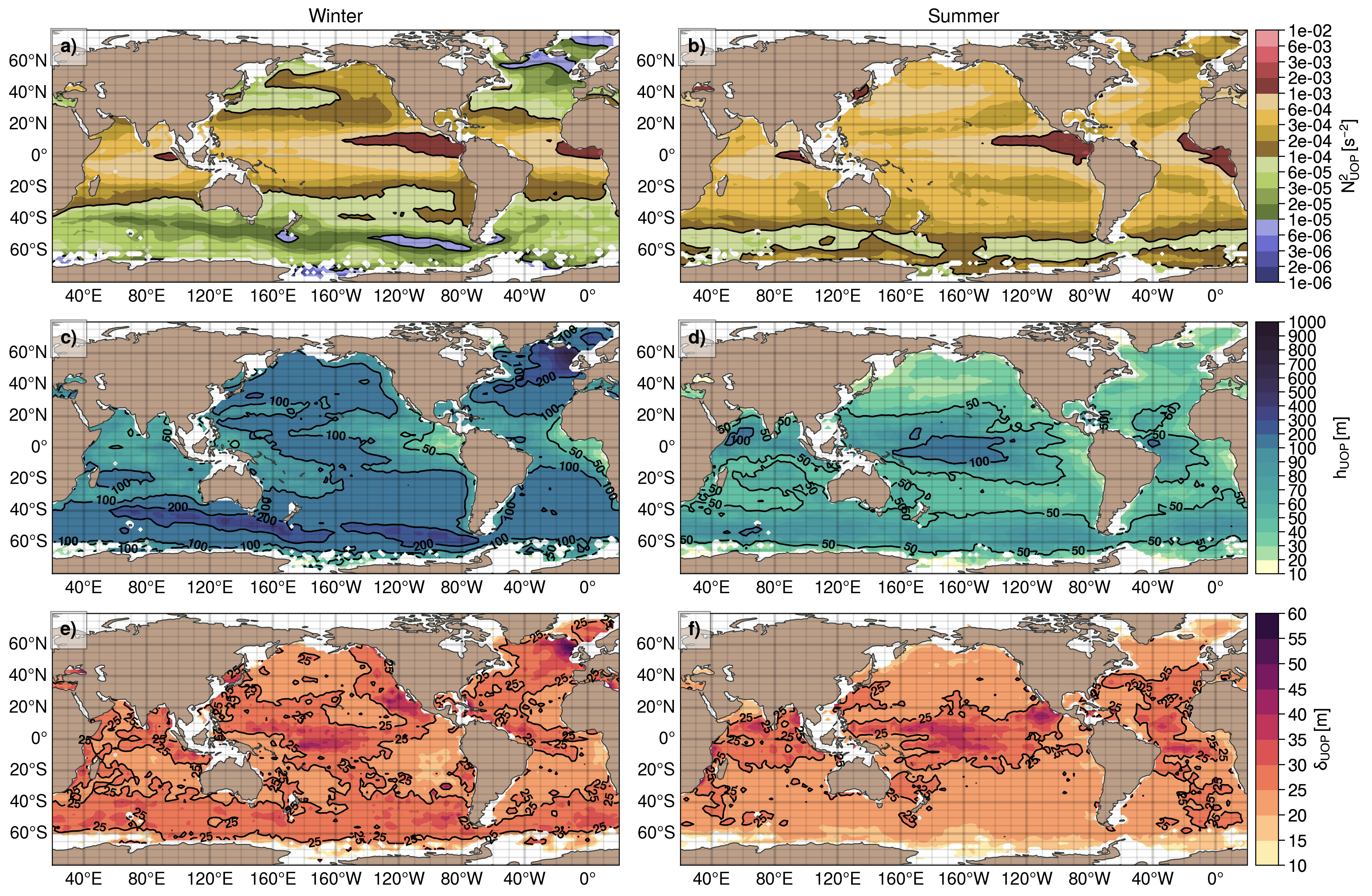}
  \end{center}
  \caption{Winter (left, a,c,e) and summer (right, b,d,f) climatologies of UOP characteristics for intensity $N^2_{UOP}$ (top, a,b), depth $h_{UOP}$ (middle, c,d), and thickness $\delta_{UOP}$ (bottom, e,f).}
  \label{fig:uop_characteristics}
\end{figure}

The map of summer UOP intensity confirms that the stratification is large and meridionally uniform north of 40\dgs, albeit with a few regions where the stratification is very large, i.e., of order $10^{-3} s^{-2}$ (Figure~\ref{fig:uop_characteristics}b). In the equatorial regions, the UOP intensity is larger on the eastern side of the basins compared to the western side. This is consistent with the classical picture of equatorial dynamics: trade winds accumulate warm waters on the western side of the basin pushing the pycnocline down, while the pycnocline on the eastern side becomes shallower and more stratified, associated with an upwelling of cold waters. The Sea of Japan is also locally more stratified in summer than the rest of the North Pacific basin. South of 40 \dgs, the summer UOP stratification remains moderate along the Antarctic Circumpolar Current $\mathcal{O}(10^{-5}\,s^{-2})$, smaller than the rest of the global ocean (Figure~\ref{fig:uop_characteristics}b). This is consistent with the deep mixing band, identified in the Southern Ocean by \cite{duvivier_argo_2018}, which is due to a right combination of factors including wind, buoyancy forcing, temperature and salinity anomalies.

Regional variability of the UOP intensity (Figure~\ref{fig:uop_characteristics}a) is more marked in winter than in summer and mostly depends on latitude (Figure~\ref{fig:meridional_variations}a). Yet, some features of the zonal variability do not show in the meridional distribution and are described here. The Kuroshio Extension and the Bering sea becomes less stratified than the rest of the basin with moderate values $\mathcal{O}(10^{-5}\,s^{-2})$. The Kuroshio Extension region indeed forms a tongue shape of moderate stratification and explains the winter intensity minimum at 40 \dgn\ in Figure~\ref{fig:meridional_variations}a. In the North Atlantic, the UOP becomes gradually less stratified with latitude north of 20\dgn\, with the weakest stratification $\mathcal{O}(10^{-6}\,s^{-2})$ reached around 55\dgn\ in the Labrador Sea, in the Irminger Sea, around the Rockall plateau and in the Norwegian Sea. Waters south of 20\dgs\ are moderately stratified $\mathcal{O}(10^{-5}\,s^{-2})$, with even smaller stratification along the ACC, reaching weak stratification values $\mathcal{O}(10^{-6}\,s^{-2})$ in the ACC west of the Drake Passage. The east-west asymmetry in the intertropical band described for the summer season remains similar in winter while the region maintains a large UOP stratification (Figure~\ref{fig:uop_characteristics}a).

\subsubsection{Depth}

The median winter UOP depth is comprised between 100 and 200 m at mid and high latitudes in winter, albeit with large variability and very deep UOPs reaching up to 600 m depth around 60\dgn\ (Figure~\ref{fig:meridional_variations}b). The UOP becomes gradually shallower in the intertropical band to reach 70--80 m at the equator. Between 10\dgs\ and 10\dgn\ , the winter UOP depth is comparable to the summer UOP depth. Outside of this band, the UOP shallows to around 30--40 m in the northern hemisphere and to around 40--50 m at midlatitudes in the southern hemisphere. The summer UOP remains deep in the Southern Ocean with values reaching 80 m, comparable to the equatorial region. 

On the map of winter UOP depth, the UOP is located between 100 and 200 m at midlatitudes over most of the Pacific ocean and the South Atlantic ocean (Figure~\ref{fig:uop_characteristics}c). UOPs deeper than 200 m are found south of 40\dgs\ along the ACC, and north of 40\dgn\ in the North Atlantic, around the North Atlantic Current. The deepest UOPs occur in the Irminger sea and around the Rockall plateau, with values exceeding 500 m, which explains the large zonal variability on the meridional distribution at these latitudes. In the equatorial Pacific, there is also an east/west asymmetry with deep UOPs on the western side (100--200 m), corresponding to the Pacific Warm Pool, and shallower UOPs on the eastern side (10-20 m). This depth asymmetry is associated with a similar asymmetry in the stratification intensity as discussed earlier, and consistent with the classical picture of Pacific equatorial dynamics. This assymmetry is also present in the equatorial Atlantic, albeit less pronounced. 

In summer, the UOP is very shallow (20--40 m) at midlatitudes in the Pacific and Atlantic oceans, but not in the midlatitude Indian ocean where there are UOPs deeper than 50 m (Figure~\ref{fig:uop_characteristics}d). This feature in the Indian ocean is probably linked to the pycnocline bowl of the Great Whirl -- a quasi-stationary anticyclonic eddy that seasonally spins up off the coast of Somalia \citep{chereskin_ekman_2002, beal_great_2013}. The summer UOP also remains deeper than 50 m in the Southern Ocean, concomitant with a weak summer UOP stratification (Figure~\ref{fig:uop_characteristics}b) and consistent with the deep mixing band described in this region \citep{duvivier_argo_2018}. At the equator, summer UOP depths and their zonal variability are similar to winter UOP depths, with values still exceeding 100 m depth in the warm pool region.

\subsubsection{Thickness}

While patterns of UOP intensity and depth exhibit substantial regional and seasonal variability, the UOP thickness is relatively constant over the global ocean with an average value of 23 m between 65 \dgs\ and 70 \dgn\ , and a standard deviation of 6.5 m. In winter, the meridional distribution of median UOP thickness oscillates between 24 and 28 m (Figure~\ref{fig:meridional_variations}c). Thicknesses larger than 35 m are rare, except at high latitudes where extreme values can reach 54-m thickness. Summer UOP thicknesses are similar to winter values from the equator to around 35$^\circ$ in both hemispheres, with values decreasing from 25-30 m at the equator to 20 m at higher latitudes. 

By looking at winter and summer maps of the UOP thickness, we can confirm that the seasonal amplitude of the UOP thickness is limited (Figure~\ref{fig:uop_characteristics}e,f). A few spots are, however, remarkable with UOPs thicker than 23 m. Areas around the Rockall plateau tend to have thicker UOP in winter reaching more than 50 m, while the UOP is thin in summer, around 20 m. In the regions of the California current and of the Pacific Warm Pool, the UOP thickness is also a bit larger than the global average, reaching 35-40 m in some places. A few spots also have a thinner UOP, with the most notable features being the region of the South Pacific Eastern North Pacific Subtropical Mode Water (SPESTMW) located around 30 \dgs\ -- 10 \dgs\ and 110\dgw\ -- 80\dgw\, and an area offshore of Angola, with UOP thicknesses between 15 and 20 meters in winter. 

Overall, the range of UOP thicknesses estimated here is consistent with the range of values estimated from local observations of the transition layer that additionally include vertical shear measurements \citep[e.g.,][]{johnston_observations_2009} and turbulence measurements \citep[e.g.,][]{kaminski_high-resolution_2021}. Note that the 5-m vertical resolution of normalised hydrographic profiles is also a limit to the accuracy of the UOP thickness.

\subsection{Amplitude and timing of the seasonal cycle}

\begin{figure}
  \begin{center}
    \includegraphics[width=\textwidth]{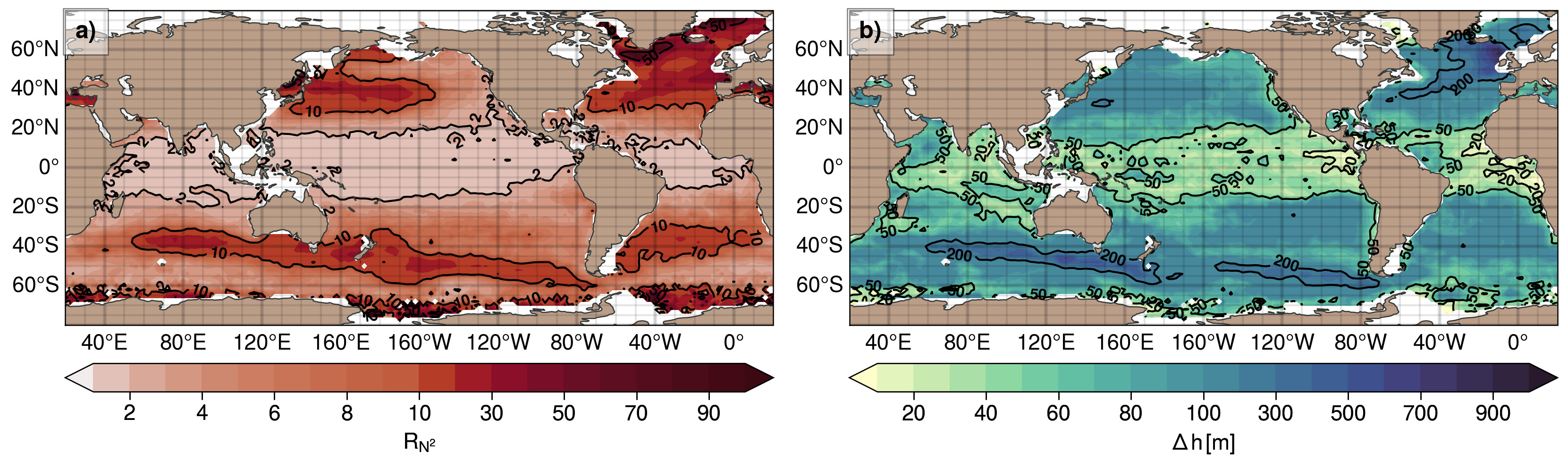}
  \end{center}
  \caption{Seasonal amplitude of UOP characteristics represented with a) the seasonal ratio of monthly maximum and minimum UOP intensity $R_{N^2}$, and b) the seasonal difference of monthly maximum and minimum UOP depth $\delta h$.}
  \label{fig:seasonal_amplitude}
\end{figure}

\begin{figure}[h!]
  \includegraphics[width=\textwidth]{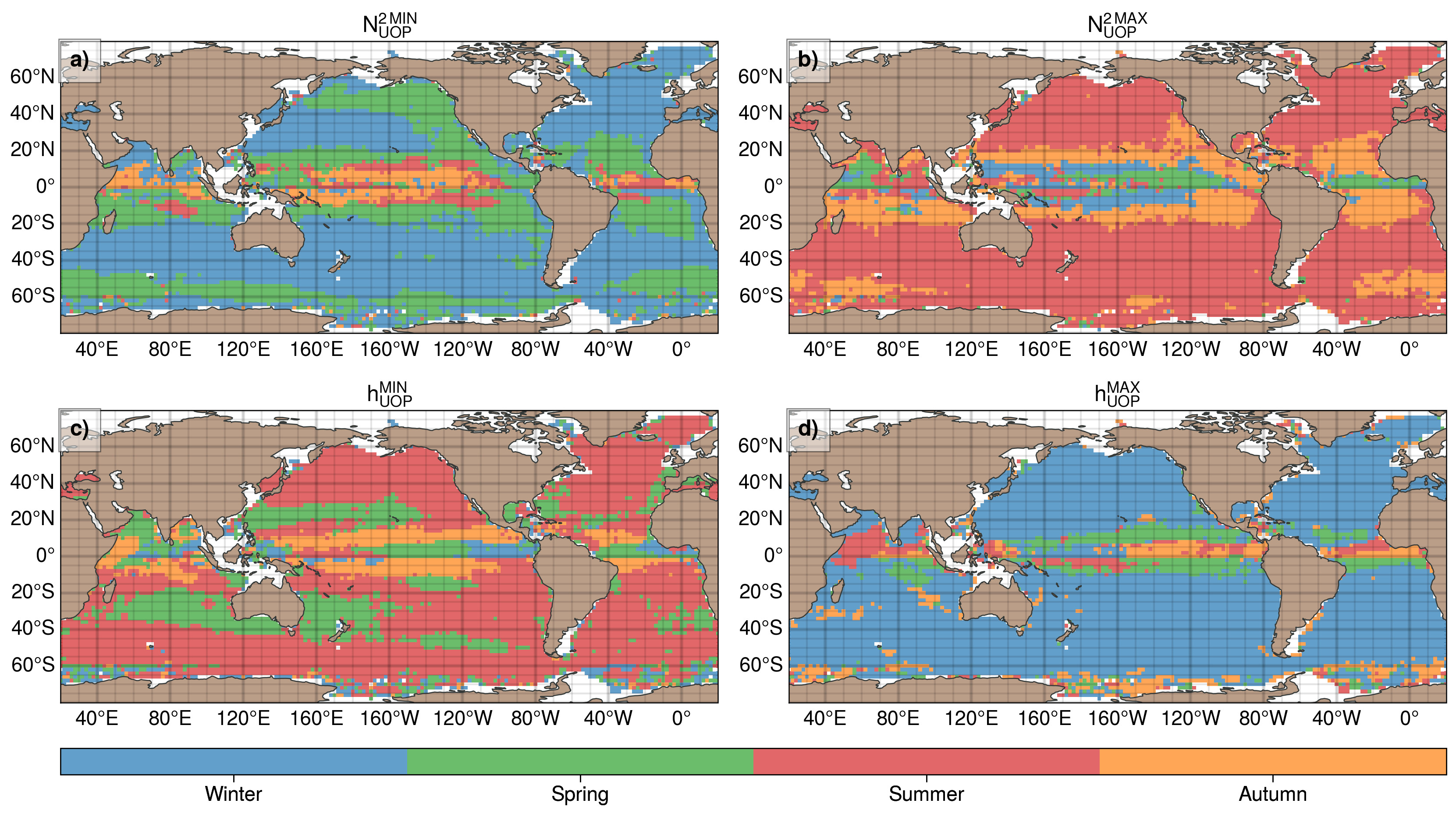}
  \caption{Seasons corresponding to a) the minimum UOP stratification, b) the maximum UOP stratification, c) the minimum UOP depth, d) the maximum UOP depth.}
  \label{fig:uop_timing}
\end{figure}

As seen earlier the UOP thickness has limited seasonal and spatial variability. We thus focus here on describing into more detail the seasonal cycle of the UOP intensity and depth. Based on the monthly maps of the UOP climatology, we found monthly maximum and minimum values for the UOP intensity and depth to evaluate the amplitude of the UOP seasonal cycle, shown in Figure~\ref{fig:seasonal_amplitude}. Along with the amplitude, we also show the corresponding seasons when the UOP characteristics are minimum and maximum in Figure~\ref{fig:uop_timing}.

At midlatitudes, some oceanic regions have a seasonal ratio between the maximum and minimum stratifications larger than 10 (Figure~\ref{fig:seasonal_amplitude}a): the Kuroshio Extension, the Bering Sea, the Sea of Japan, the North Atlantic basin north of 30 \dgn, the Mediterranean Sea, and some sections of the ACC. While seasonality is important over the full width of the North Atlantic basin, most of the seasonality is concentrated on the western side of the North Pacific basin. The seasonal ratio of the UOP intensity is larger than 50 in the Labrador sea and in the Nordic seas. These regions have thus the strongest seasonality in the global ocean. Within the subtropical gyres, the amplitude of the UOP depth exceeds 100 m. It becomes greater than 200 m in some regions of the ACC and around the North Atlantic Current (Figure~\ref{fig:seasonal_amplitude}b). The largest depth amplitude occurs in the region of the Irminger Sea and around the Rockall plateau. Outside the intertropical band, the minimum UOP stratification occurs in winter, except in the northern and eastern part of the North Pacific, and in some places of the ACC, where it occurs later in spring (Figure~\ref{fig:uop_timing}a). In this part of the global ocean, the winter season also corresponds to the maximum depth of the UOP (Figure~\ref{fig:uop_timing}d). The maximum UOP stratification occurs generally in summer outside the intertropical band with a few exceptions in the Atlantic part of the ACC (Figure~\ref{fig:uop_timing}b). The summer season also corresponds to the shallowest UOPs, with a few regions where it occurs earlier in spring (i.e., Agulhas leakage, East Australian Current region, the Indian midlatitudes, and the North Pacific Subtropical Countercurrent region). 

At midlatitudes of the northern hemisphere, the timing of the UOP intensity is consistent with the minimum and the maximum of sea surface temperature and the stratification index used by \cite{somavilla_warmer_2017}, that occurs respectively at the end of winter and at the end of summer. The timing of the maximum MLD estimated by \cite{somavilla_warmer_2017} and by \cite{johnson_gosml_2022} also agrees with the maximum UOP depth occurring at the end of winter. However, while the minimum of the MLD estimate used by \cite{somavilla_warmer_2017} also occurs at the end of winter, we find that the minimum UOP depth occurs in summer, consistently with \cite{johnson_gosml_2022} who found a similar timing for the minimum of their MLD estimate. This difference with \cite{somavilla_warmer_2017} is possibly due to their method used for estimating the MLD \citep{gonzalez-pola_vertical_2007}.

In the intertropical band (20\dgs -- 20 \dgn), the seasonality is different from the other latitudes. The seasonal ratio of the maximum and the minimum UOP stratification is small, with most of the regions having a ratio smaller than 2 (Figure~\ref{fig:seasonal_amplitude}a). The seasonal amplitude of the UOP depth is also limited and less than 100 m, with most of the regions having a seasonal amplitude less than 50 m (Figure~\ref{fig:seasonal_amplitude}b). The smallest amplitude is found close the western boundaries of the Pacific and Atlantic equatorial ocean, with values less than 20 m. The seasons of the maximum and minimum UOP characteristics is more complex in the equatorial region, though some patterns seems to emerge (Figure~\ref{fig:uop_timing}). A minimum stratification may be noticed close to the equator in summer and in autumn, particularly in the Pacific ocean, while it mostly occurs in spring closer to the tropics (Figure~\ref{fig:uop_timing}a). The reverse behaviour is similar for the maximum intensity, occurring in winter and spring close to the equator, and in autumn closer to the tropics (Figure~\ref{fig:uop_timing}b). Between 0$^\circ$ and 5\dgn\, the minimum UOP depth tends to occur in winter and spring in the Pacific and Atlantic basins (Figure~\ref{fig:uop_timing}c). Between the equator and the tropics, the minimum depth rather occurs in autumn and sometimes in summer. As for the UOP intensity, the seasonality is reversed for the maximum UOP depth, occurring mainly in autumn and summer between 0$^\circ$ and 5\dgn\, and in spring in the rest of the 10\dgs --10 \dgn\ band. Note that the tropical Indian ocean is apart and has an even more complex UOP timing than the other basins, linked to the specific regime of the reversing moonson in this area \citep{schott_indian_2009} and to the seasonality of the Great Whirl \citep{beal_great_2013}. Thus, the complex UOP seasonality in the intertropical band is to be investigated in the future, but should be partly explained by precipitation regimes occurring in this region.

\subsection{Intermediate seasons and restratification}

We now focus on intermediate seasons using the upper ocean stratification index UOSI, which corresponds to the stratification index $SI(z)$ of equation~\ref{eq:si_n2} estimated at the UOP lower limit ($h_{UOP}^{-}$). As shown in section~\ref{sec:data_method}, the variations of this quantity may be linked to the intensity of buoyancy forcing \citep[e.g.,][]{herrmann_modeling_2008}. Thus, the stratification index UOSI may be useful to capture restratification periods due to the ease of adverse surface buoyancy fluxes or to the relative importance of lateral fluxes, including those associated with submesoscale motions \cite[e.g.][]{johnson_global_2016, du_plessis_submesoscale_2017}.

\begin{figure}[h!]
  \includegraphics[width=\textwidth]{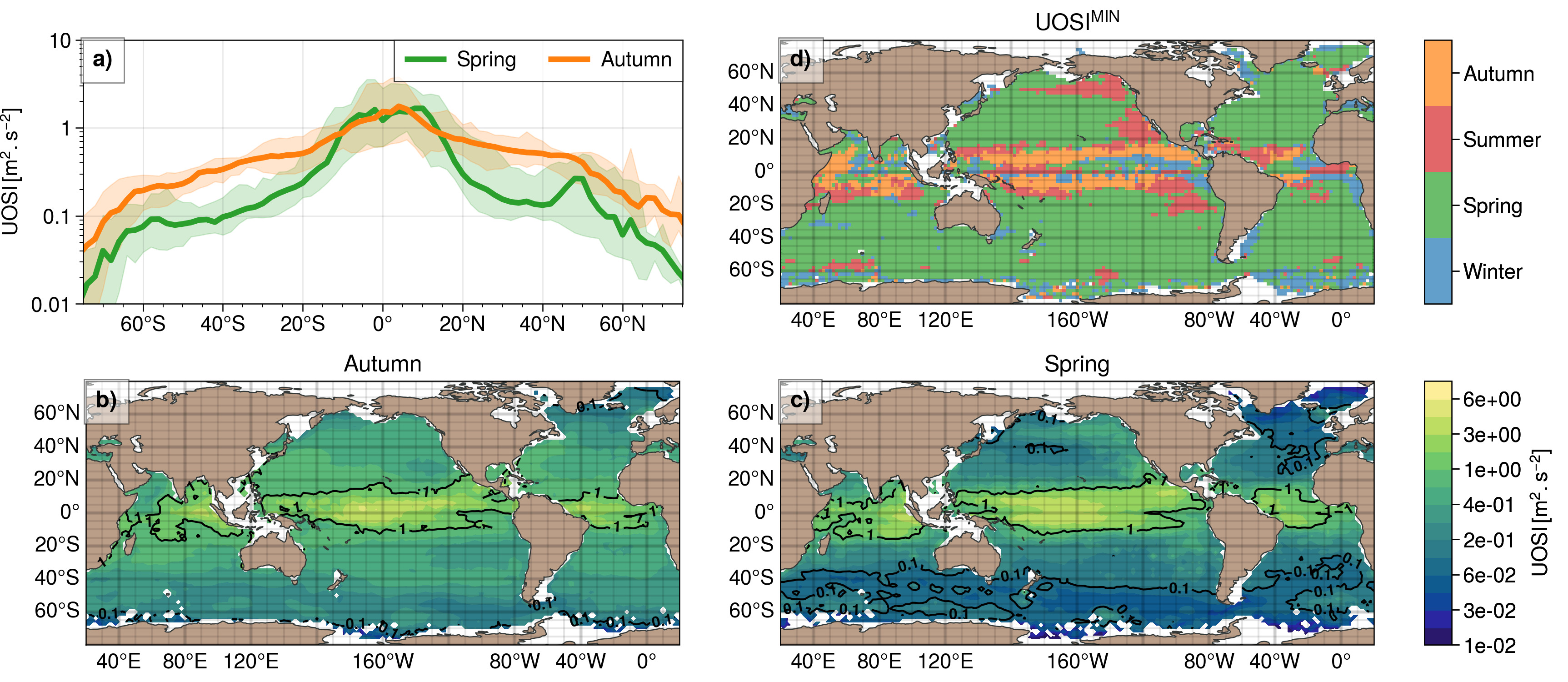}
  \caption{a) Meridional distribution of median UOSI for the autumn season (orange curve) and the spring season (green curve). The shaded areas represent the amplitude of the distributions based on the 5 and 95 percentiles. b) Seasons corresponding to the minimum UOP stratification index. c,d) Autumn and spring climatologies of UOP stratification index.} 
  \label{fig:uop_bc}
\end{figure}

The largest UOSI occurs around the equator, between 10\dgs\ and 10\dgn , where the UOP is the most difficult to erode due to large stratification values (Figure~\ref{fig:uop_bc}a). In these low-latitude regions, the UOSI seasonality is very limited and values are larger than 1 $m^2.s^{-2}$ on the meridional distribution (Figure~\ref{fig:uop_bc}a). These large UOSI values concern most areas near the equator as shown by autumn and spring maps (Figure~\ref{fig:uop_bc}c,d), and suggest that exchanges between the upper ocean and the interior ocean are very limited all time of the year in these regions. 

The UOSI decreases with latitude, with a larger drop in spring between 10$^\circ$ and 20 $^\circ$ (Figure~\ref{fig:uop_bc}a). The minimum in UOSI occurs in spring at latitudes higher than 15--20$^\circ$, consistent with a newly stratified layer forming near the ocean surface during this season, that can be easily eroded by external buoyancy loss (Figure~\ref{fig:uop_bc}b). Latitudes higher than 40$^\circ$ in the Southern Ocean and in the North Pacific tend to have the smallest UOSI in spring, with values smaller than 0.1 $m^2.s^{-2}$ (Figure~\ref{fig:uop_bc}d). In spring, the eastern and northern sides of the North Pacific remains with UOSI values comparable to those in autumn, while the rest of the basin has reduced values (Figure~\ref{fig:uop_bc}c,d). This corresponds to a bump at 50\dgn\ on the meridional distribution (Figure~\ref{fig:uop_bc}a) and is linked to a minimum UOSI reached latter in summer in these regions (Figure~\ref{fig:uop_bc}b).

\subsection{Variability}

\begin{figure}[h!] 
 \includegraphics[width=\textwidth]{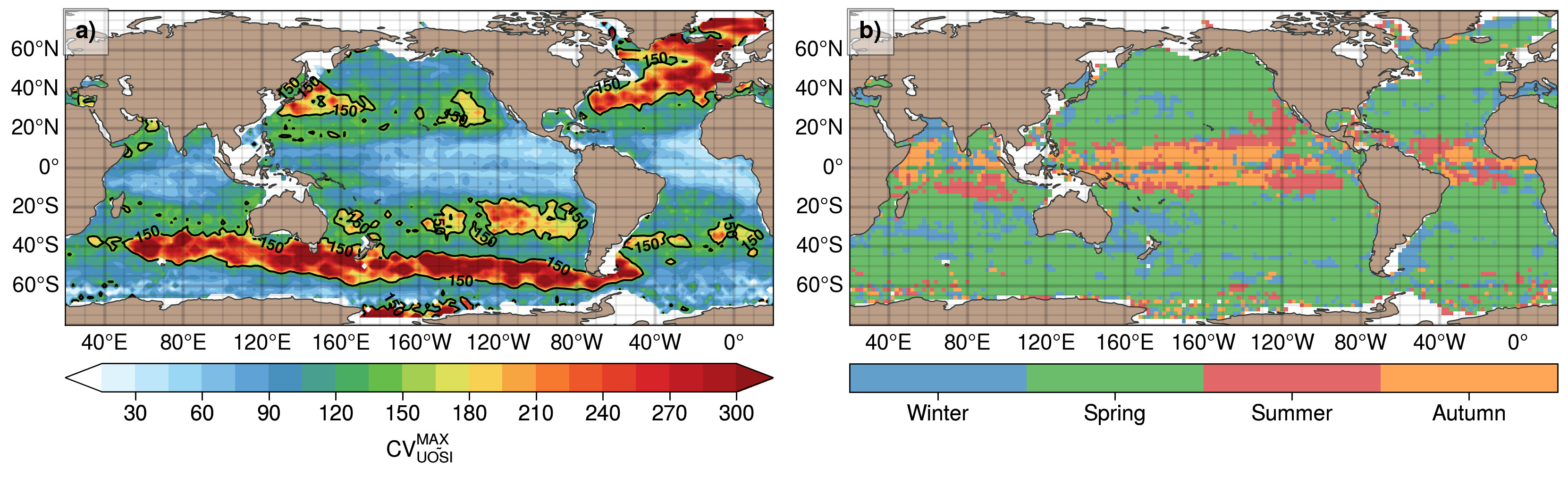}
  \caption{a) Maximum coefficient of variation for the stratification index UOSI and b) the corresponding season.}
  \label{fig:uop_variability}
\end{figure}

Since the UOSI integrates information on the depth, intensity and thickness of the UOP, we focus on the variation of this former quantity. The variability described here includes both sub-monthly and interannual variability as well as spatial variability with scales smaller than the bin size (i.e., 2 $^\circ$). We evaluate the relative variability of UOP characteristics for each month and each spatial bin using the coefficient of variation $CV$, that is simply the ratio of the standard deviation and the mean. Results for the maximum $CV$ and the corresponding season are shown in Figure~\ref{fig:uop_variability}. The variability characterised here includes both sub-monthly variability and interannual variability. Coefficient of variations are also given for the UOP characteristics --intensity, depth, and thickness -- in Supplementary Materials, to complement the UOSI coefficient of variation. 

Regions of large relative variability for the UOSI are similar to region with large UOP seasonal amplitude  (Figure~\ref{fig:seasonal_amplitude}a), with the largest variability occurring north of 40 \dgn\ in the North Atlantic basin and along the Indian and Pacific part of the ACC (Figure~\ref{fig:uop_variability}). Large variability is also noticed to a lesser extent in the Kuroshio Extension, in the regions of the Pacific subtropical mode waters offshore of California and offshore of Chile, in the Agulhas leakage region, and in the Coral Sea. In all these regions, the coefficient of variation $CV$ exceeds 150\%. The maximum relative variability is generally reached in spring for latitudes higher than 10--15 $^\circ$, with a few regions having the maximum occurring in winter (Figure~\ref{fig:uop_variability}b). In the intertropical band, between 15 \dgs\ and 15 \dgn , the coefficient of variation is weaker especially in the eastern side of the basins with values smaller than 50 \% (Figure~\ref{fig:uop_variability}a). This minimum in variability generally occurs in autumn and summer (Figure~\ref{fig:uop_variability}b). Note that the regions with large CV in UOSI as described above tend to correspond with spring MLD distributions being skewed towards deeper values and having stronger tails than a normal distribution \citep[see Figures 8 and 9 in][]{johnson_gosml_2022}.

\section{Regional analyses of the UOP}\label{sec:regional_analyses}

We have shown that the different UOP characteristics for the summer and winter seasons only, and pinpointed some particularities of certain regions of the world's ocean. In this section, we focus into more detail on the seasonal cycle of certain regions, most of them having substantial variability of the UOSI (i.e., $CV \geq 150 \%$ in Figure~\ref{fig:uop_variability}a): Kuroshio Extension (KE, 140\dge --170 \dge, 25\dgn --40 \dgn), North Pacific Eastern Subtropical Mode Water (NPESTMW, 145\dgw --120\dgw, 20\dgn --40\dgn), Gulf Stream (GS, 70\dgw --40\dgw, 30\dgn --45\dgn), Labrador Sea (60\dgw --40\dgw, 55\dgn --65\dgn), Rockall Plateau (RP, 20\dgw --5\dgw, 50\dgn --65\dgn), Indian ACC (IACC, 80\dge --120\dge, 50\dgs --40\dgs), South Pacific Eastern Subtropical Mode Water (SPESMW, 120\dge --80\dge, 35\dgs --20\dgs), Agulhas Leakage (AL, 0$^\circ$--20\dgn, 45\dgs --25\dgs). We also add the Pacific Warm Pool (PWP, 180$^\circ$--140\dgw, 10\dgs --5\dgn) to the study because this region contrasts with larger latitudes and it has a moderately deep UOP at around 100 m (Figure~\ref{fig:uop_characteristics}c). Whilst the global climatology showed mainly the UOP median values computed in 2$^\circ\times$ 2$^\circ$ bins, regional statistics are computed in this section by taking all the hydrographic profiles within each of the regions. In these regions, we found that February and August are well representative of the minimum and maximum positions of the UOP for the summer and winter seasons, so that these months will be used to illustrate some of our results. We also added information on the permanent ocean pycnocline (POP) depth using the dataset produced by \cite{feucher_subtropical_2019}. This dataset will help to discuss the relative position of the UOP and the POP.

\subsection{Stratification profiles}\label{sec:regional_profiles}

\begin{figure}[h!]
  \includegraphics[width=\textwidth]{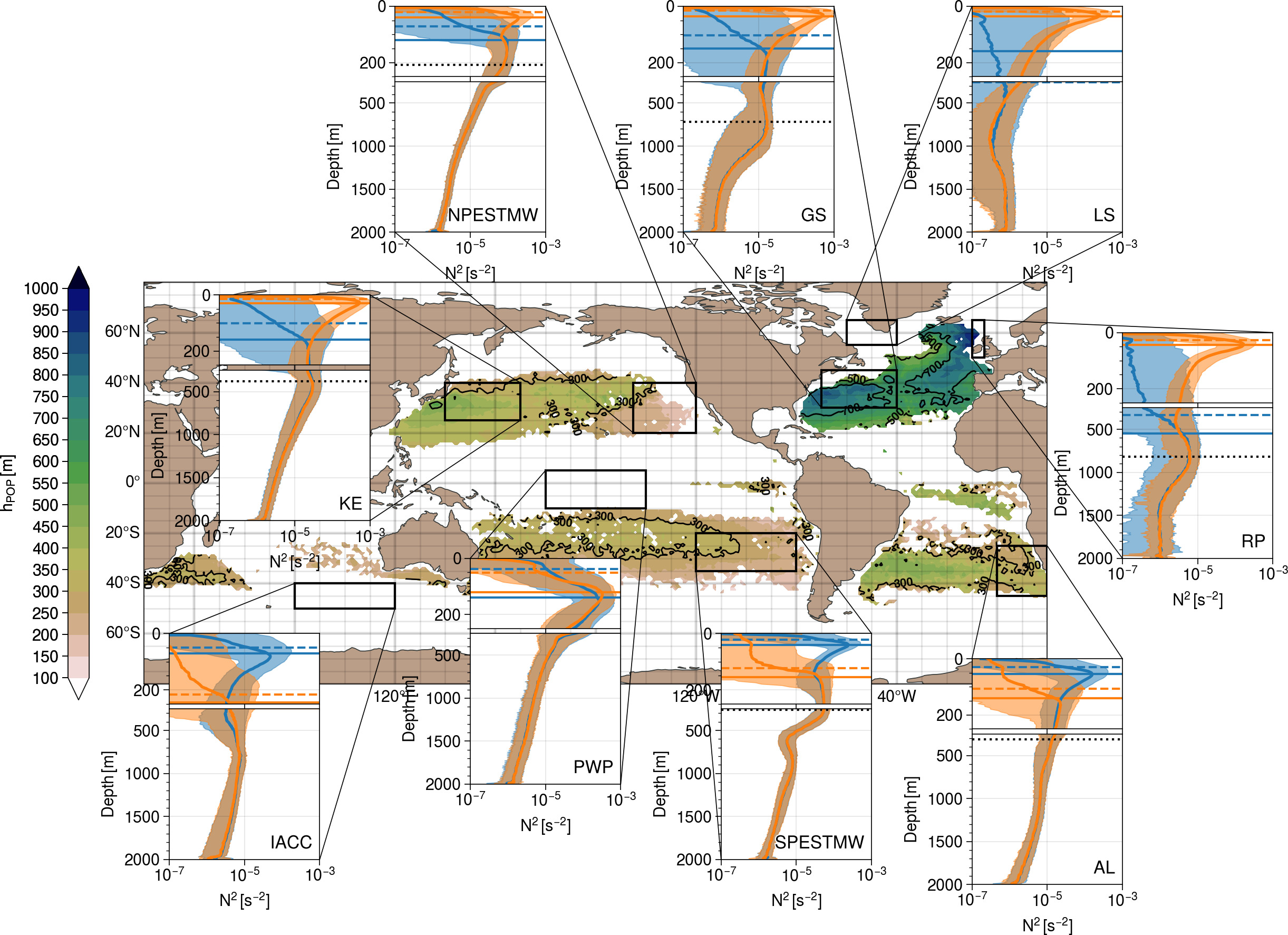}
  \caption{Depth of the subtropical permanent ocean pycnocline $h_{POP}$ reproduced from \cite{feucher_subtropical_2019}'s dataset. Median profile of stratification $N^2$ and the envelop plotted using the first and second quartiles are added for February (blue curves) and August (orange curves) for nine regions of study: 
  Kuroshio Extension (140\dge --170 \dge, 25\dgn --40 \dgn),
  North Pacific Eastern Subtropical Mode Water (145\dgw --120\dgw, 20\dgn --40\dgn),
  Gulf Stream (70\dgw --40\dgw, 30\dgn --45\dgn),
  Labrador Sea (60\dgw --40\dgw, 55\dgn --65\dgn),
  Rockall plateau (20\dgw --5\dgw, 50\dgn --65\dgn),
  Indian ACC (80\dge --120\dge, 50\dgs --40\dgs), 
  Pacific Warm Pool (180$^\circ$--140\dgw, 10\dgs --5\dgn)
  South Pacific Eastern Subtropical Mode Water (120\dge --80\dge, 35\dgs --20\dgs),
  Agulhas Leakage (0$^\circ$--20\dgn, 45\dgs --25\dgs). The UOP depth and the MLD $h_{MLD}^{MIN}$ are represented for each season with plain and dashed horizontal lines, respectively. The median depth of the permanent pycnocline within each region is added with black dotted lines on the profile panels.}
  \label{fig:profile_regions}
\end{figure}

The climatological map of the permanent ocean pycnocline (POP) is shown in Figure~\ref{fig:profile_regions}, along with the location of the aforementioned regions of study and the corresponding stratification profiles $N^2(z)$ in February and in August. Consistent with results from the previous section, the median $N^2$ profiles confirm that upper ocean stratification varies seasonally outside the tropics with substantial variations associated with the UOP (blue and orange plain lines in stratification panels of Figure~\ref{fig:profile_regions}). The summer stratification peak associated with the UOP is large and sharp, while the winter stratification peak is much less marked on average.  The upper ocean stratification is also much less variable near the equator, as is the UOP (e.g., PWP profiles). In the PWP region, a first stratification variation occurs near the surface around and is probably linked to salinity barrier layers \citep[e.g.,][]{lukas_mixed_1991, vialard_ogcm_1998}. This density variation is related to a subsurface halocline that is distinct and shallower than the thermocline. While this layer imprints on certain MLD variables \citep{de_boyer_montegut_control_2007, mignot_control_2007}, it is not captured by our UOP method, either because of the absence of a clear stratification peak or because its amplitude is much weaker than the stratification peak associated with the permanent pycnocline.

In most of the regions plotted, the median winter UOP is located where the winter and summer stratification profiles start to differ (KE, NPESTMW, GS, RP,  SPESTMW, AL). In the two Pacific regions where subtropical mode waters are formed (i.e., NPESTMW and SPESTMW), the mode water signatures are clear with a local minimum in stratification located between the summer UOP and POP. In winter, this minimum disappears suggesting that the mode water layer may be influenced by air-sea exchanges.

The POP values taken from the climatology of \cite{feucher_subtropical_2019}, shown with black dotted lines in Figure~\ref{fig:profile_regions}, seem consistent with our dataset since the POP is located on a local maximum of the median stratification profile (i.e., KE, NPESTMW, GS, RP, SPESTMW). In these regions, the winter UOP is always shallower than the POP, suggesting that the UOP does not always merge with the POP in winter. In all the regions of study, the UOP is located below the MLD (blue and orange dash lines), except in winter in the Labrador Sea (LS). The depth separating the MLD and the UOP is usually small in summer, but it may slightly increase in winter (e.g., KE, NPESTMW, AL). In the Labrador Sea, the median winter MLD ($\sim 250$ m) is found to be much deeper than the median winter UOP depth ($\sim 150$ m). This inverse behaviour could be due to a sensitivity of our method to very small stratification peaks in a region with very small stratification values over the 0--2000 m depth range. We will discuss later the relative position of the MLD and the upper limit of the UOP, along with the validity of our peak detection method.

\subsection{Seasonal cycle of UOP intensity and depth}

To complement the analysis made with the stratification profiles that showed only two months of the year of the median UOP depth, we show the full monthly seasonal cycle of the median UOP in the intensity-depth space, plotted on a log-log scale for the different regions of study (Figure~\ref{fig:2d_distribution}). In some regions, the seasonal cycle in the intensity-depth space looks regular and describes an ellipsoid trajectory (Figure~\ref{fig:2d_distribution}a,c,d,i). In other regions, depth changes during restratification times in spring occurs more abruptly than stratification changes, yielding instead a D-shape (Figure~\ref{fig:2d_distribution}b,e,f,h). The seasonal cycle is even more distorted around the Rockall plateau (RP) where the UOP deepening also occurs more abruptly than destratification during late autumn and early winter (Figure~\ref{fig:2d_distribution}b). As shown earlier for the global climatology, the seasonal cycle in the Pacific Warm Pool (PWP) is very limited (Figure~\ref{fig:2d_distribution}f). 

These results suggest that the seasonal evolution of the UOP does not occur at the same pace everywhere in the global ocean. This is consistent with the variety of MLD seasonal cycles showed by \cite{johnson_gosml_2022}, with sometimes abrupt or more gradual rise of the MLD (see their Figure~2). Note that the monthly median UOPs are generally concentrated in distinct areas of the intensity-depth space for winter and summer, while months of intermediate seasons connect these summer and winter areas.

\subsection{Winter and summer UOP variability}

\begin{figure}[h!]
  \includegraphics[width=\textwidth]{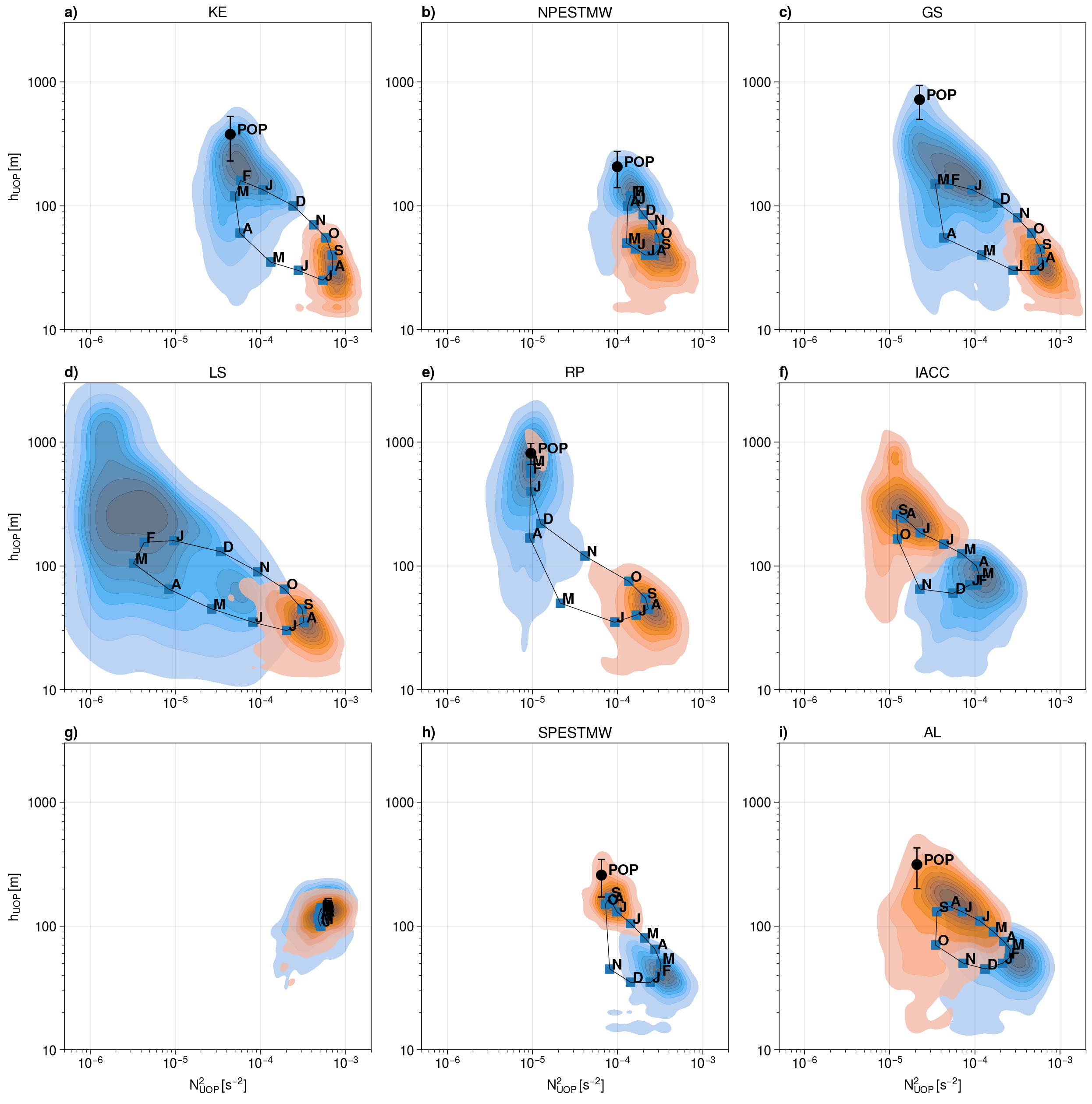}
  \caption{Seasonal cycle of the median represented in the logarithmic intensity-depth space with squares labelled with the first letter of the month. Joint distributions of intensity and depth are added for February (blue shading) and August (orange shading).}
  \label{fig:2d_distribution}
\end{figure}

The regional and temporal variability of February and August UOPs are analysed with intensity-depth paired distribution shown in Figure~\ref{fig:2d_distribution}. In general, the amplitude ratio (i.e., the maximum or the minimum) of the UOP variability is larger in winter than in summer. The winter UOP has only moderate variability in the Pacific mode water regions (NPESTMW, SPESTMW, Figure~\ref{fig:2d_distribution}b,h), of the same order as in summer, and less than in the other regions in winter at mid and high latitudes (e.g., KE, GS, LS, RP, IACC, AL). In the two major western boundary currents of the northern hemisphere, the depth variations are important in winter with the UOP depth being as shallow as 20 m and possibly reaching 700 m and 1000 m in the Kuroshio Extension (KE) and in the Gulf Stream (GS), respectively (Figure~\ref{fig:2d_distribution}a,c). The stratification however remains weak to moderate in winter ($10^{-5}$ to  $3.10^{-4} s^{-2}$). A similar picture is found in the regions of the Indian ACC (IACC) and of the Agulhas Leakage (AL) for the winter month (Figure~\ref{fig:2d_distribution}f,i). Around the Rockall plateau the winter UOP intensity is concentrated around weak values of stratification without much variability $\mathcal{O}(10^{-5} s^{-2})$ (Figure~\ref{fig:2d_distribution}e). However, the depth variability is much more pronounced in winter spanning a range from 40 m to 2000 m. In winter, the Labrador Sea (LS) shows the widest distribution both in stratification and depth, as well as the deepest and least stratified UOPs (Figure~\ref{fig:2d_distribution}d). The distribution spans a range of 20 m to 2000 m in depth and a range of $5.10^{-6}\,s^{-2}$ to $10^{-4}\,s^{-2}$ in intensity, consistent with deep MLDs observed in this region \cite[e.g.][]{johnson_gosml_2022}.

Interestingly, it is possible to find moderately deep UOP in summer at mid and high lattitudes that may reach 80 m to 100 m depth (e.g., NPESTMW, RP, SPESTMW, AL), and even more in the Indian ACC (Figure~\ref{fig:2d_distribution}f). This result is consistent with a recent statistical monthly climatology of the global ocean surface mixed layer \citep{johnson_gosml_2022}, that shows distribution tails of the MLD reaching up to 100 m in late summer. The UOP stratification generally remains larger than $10^{-4} s^{-2}$ in summer, except in regions involving the ACC (IACC and AL). This is consistent with the summer climatology of UOP intensity (Figure~\ref{fig:uop_characteristics}b), showing moderately stratified values $\mathcal{O}(10^{-5} s^{-2})$ along the ACC. Finally, the variability is similar between August and February in the Pacific Warm Pool (PWP), of the same order as summer variability in the other regions, and consistent with a weak seasonal cycle of the median in this region. Note that both winter and summer UOP distributions may partially overlap in certain regions (NPESTMW, GS, LS, IACC, AL). The depth variations of the UOP described here agree with the statistical monthly climatology of the MLD produced by \cite{johnson_gosml_2022}, who showed much wider MLD distributions in winter than in summer (see their Figure~5 in particular).
 
The location of the POP and its thickness are represented in the intensity-depth space in Figure~\ref{fig:profile_regions}. In all the regions studied where the POP is available, the winter UOP paired distribution overlaps with the POP (Figure~\ref{fig:2d_distribution}a,b,c,e,h,g). This gives confidence in the consistency between our method, that finds the UOP (corresponding sometimes to the POP during winter months), and the method developed by \cite{feucher_subtropical_2019}, that finds the POP. In some regions (i.e., GS, SPESTMW, AL), the POP is located on the distribution tail, which may suggest that the UOP does not reach the POP on average but merging between the UOP and the POP could rather happen intermittently (in space and time). In the other regions, and especially around the Rockall Plateau, the POP is closer to the center of the UOP distribution, suggesting that the UOP is more likely to merge with the POP.

\subsection{UOSI seasonal cycle and variability}

\begin{figure}[h!]
  \includegraphics[width=\textwidth]{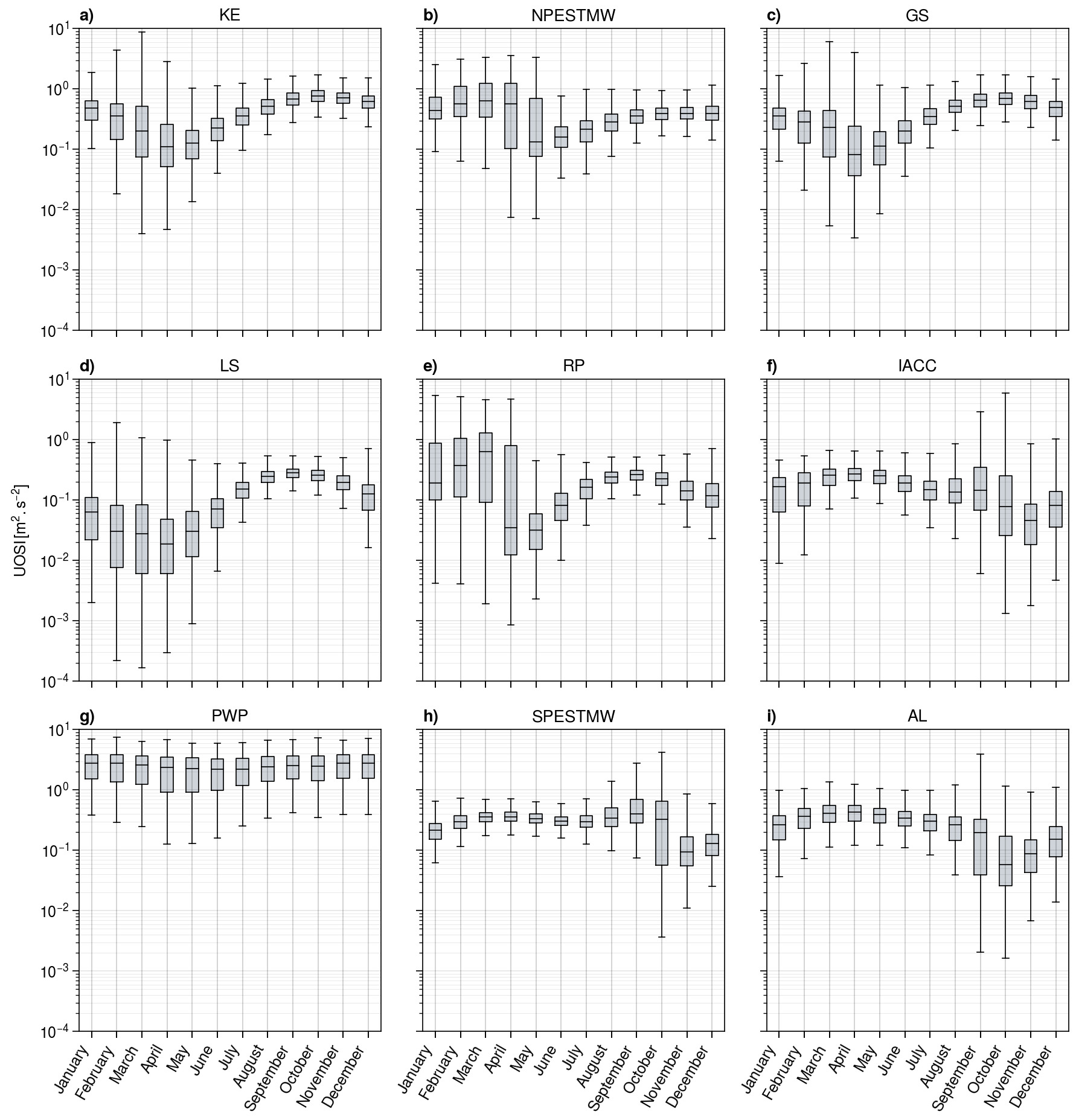}
  \caption{Seasonal cycle of the upper ocean stratification index (UOSI) distributions for the different region of analysis. The boxes represent the first and third quartiles, with the median drawn within. The whiskers are drawn within the 1.5 interquartile range.}
  \label{fig:boxplots_buoyancy_content}
\end{figure}

The seasonal cycle of upper ocean stratification statistics is investigated in each region using the monthly distributions of the UOSI, illustrated with boxplots in Figure~\ref{fig:boxplots_buoyancy_content}. Similar seasonal cycles for the UOP depth and intensity are also given in Supplementary Materials. The eddy active regions studied at mid latitudes (i.e., KE, GS, IACC, AL) see their median UOSI gradually decreasing throughout autumn and winter to reach a minimum in April in the northern hemisphere regions (Figure~\ref{fig:boxplots_buoyancy_content}a,c) and in October/November in the southern hemisphere regions (Figure~\ref{fig:boxplots_buoyancy_content}f,i). Note that this gradual decrease is halted in the Agulhas Leakage (AL) region between September and October with a notable drop in UOSI. According to equation~\ref{eq:dsi_dt}, a negative variation in UOSI in winter means that the upper ocean layer experiences adverse buoyancy fluxes during this season, i.e. large winds and cooling. The Labrador Sea (LS) shows a similar seasonal pattern (Figure~\ref{fig:boxplots_buoyancy_content}d), with the specificity of having amongst the smallest values of UOSI in winter. This small UOSI means that the upper ocean stratification may be easily eroded during this time of the year in this region. The gradual decrease in UOSI is consistent with the ellipsoid trajectory of the seasonal cycle in the intensity-depth space for the regions KE, GS, LS and AL (Figure~\ref{fig:2d_distribution}a,c,d,i).

In the region of the Pacific mode waters (NPESTMW, SPESTMW) and around the Rockall plateau (RP), the upper ocean layer gains buoyancy (i.e., positive fluxes) throughout winter with an increase of the UOSI after a first decrease in autumn. In this case, the UOSI increases because the effect of the deepening of the UOP during wintertime counteracts the effect of the weakening in UOP stratification (see the depth and intensity boxplots in Supplementary Materials). A substantial jump in UOSI occurs in these regions in early spring, depicting the rapid formation of a weakly stratified layer near the surface, with the possibility of burying a fossil layer just below. This is associated with a jump to shallower UOP depths during this same month and the D-shape trajectory of the seasonal cycle in the intensity-depth space (Figure~\ref{fig:2d_distribution}b,e,h, see also the UOP depth boxplots in Supplementary Materials). 

All the regions show a gradual increase in UOSI throughout spring and summer after the minimum value has been reached in spring, meaning that the rise and the intensification of the UOP occur smoothly after restratification has taken place. The timing of the minimum UOSI is also consistent with the maps shown in Figure~\ref{fig:uop_bc}d and with the stratification index of \cite{somavilla_warmer_2017} studied in a few midlatitude regions.

The UOSI variability generally intensifies throughout winter and reaches its maximum in early spring (Figure~\ref{fig:boxplots_buoyancy_content}), consistently with the map of the timing of maximum relative variability (Figure~\ref{fig:uop_bc}d). In some regions the UOSI variability during spring restratification is much larger than the variability levels noticed for the rest of the year (SPESTM, NPESTMW, AL). The upper ocean restratification is generally followed by a reduction in UOSI variability as the newly formed layer near the surface stabilises in time and space. In the Pacific Warm Pool (PWP), the seasonality of the median UOSI is small but the variability levels remain substantial all year long, with amplitude ratio comprised between those noticed for summer and winter at higher latitudes. In this region, no restratification phase may be noticed as the UOSI remains large. 

Overall, this regional study shows that the seasonal cycle of the UOP in the intensity-depth space and the process of restratification are not homogeneous over the global ocean with different behaviours. In particular, restratification occurs when a minimum in UOSI is reached, generally associated with a large variability whose amplitude depends on the region. Restratification may sometimes occurs gradually or more abruptly depending on the region as shown by the shape of the seasonal cycle in the intensity-depth space (Figure~\ref{fig:2d_distribution}) and by the seasonal distribution of the UOSI (Figure~\ref{fig:uop_bc}).

\section{Discussion}\label{sec:discussion}

\subsection{UOP and MLD estimates}

In order to evaluate the capacity of our UOP definition to be consistent with the position of the mixed layer, we compared the location of the UOP upper boundary $h_{UOP}^{+}$ relative to the minimum of classical MLD estimates $h_{MLD}^{MIN}$, as defined in section~\ref{sec:data_method}.

\begin{figure}[h!]
  \begin{center}
    \includegraphics[width=15cm]{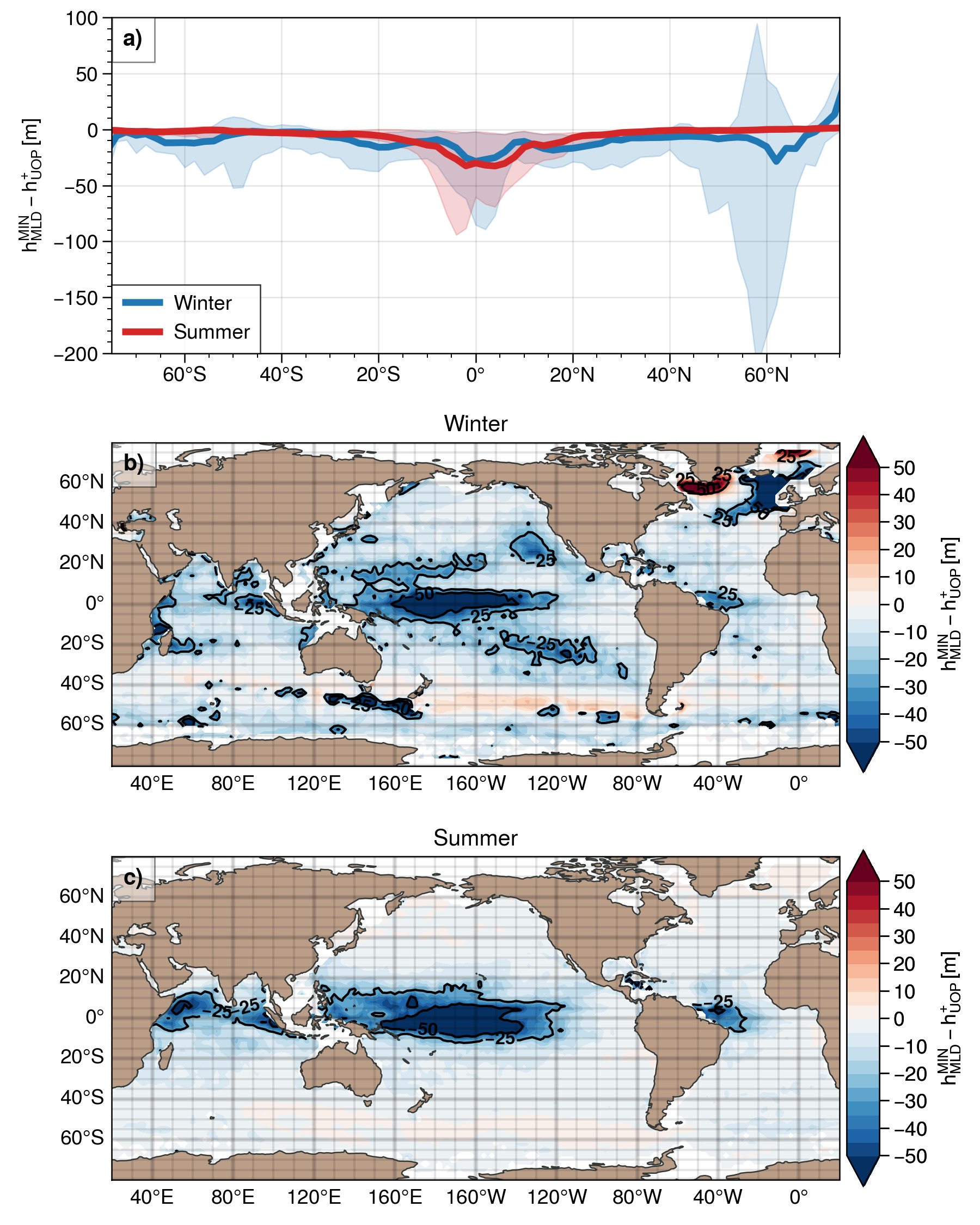}
  \end{center}
  \caption{Median of the difference between the minimum of mixed layer depth estimates ($h_{MLD}^{MIN}$) and the upper boundary of the UOP ($h^{+}_{UOP}$). a) Meridional distribution of the median difference for the winter season (blue curves) and for the summer season (red curves). The shaded areas represent the amplitude of the distributions based on the 5 and 95 percentiles. b), c) Climatological maps of winter and summer median differences.}
\label{fig:MLD_vs_UOP}
\end{figure}

On the zonal distribution, $h_{UOP}^{+}$ and $h_{MLD}^{MIN}$ are very close outside the intertropical band (15\dgs --15\dgn), with very little spread on the distribution for the summer season. Near the equator, the difference is however significant with the full distribution showing that $h_{UOP}^{+}$ is deeper than $h_{MLD}^{MIN}$ (around 30 m deeper for the median). The largest differences noticed near the equator on the meridional distribution may reach 90 m and are explained by large differences in the Pacific Warm Pool region seen both on summer and winter climatological maps (Figure~\ref{fig:MLD_vs_UOP}b,c). As already discussed in section~\ref{sec:regional_analyses}, this is explained by the salinity barrier layers having an imprint on the stratification that is too small to be captured by our UOP method.

In winter, the high latitudes have larger differences between $h_{UOP}^{+}$ and $h_{MLD}^{MIN}$ with a large dispersion at around 55--60 \dgn. This is mainly due to the UOP upper boundary being shallower than the minimum of the MLD estimates in the Labrador Sea and in the Nordic Seas ($< 25$ m). Although the absolute difference is substantial in these regions, the relative difference represents less than 20\% (see Supplementary Materials) so that one may suggest that the correspondence between the UOP and the minimum of the MLD estimate is actually acceptable. Much deeper UOP than MLD estimates ($> 50$ m) occurs around the Rockall Plateau and along the ACC south of Tasmania and New Zealand. This is due to the presence of vertically compensated layers in these regions \citep{de_boyer_montegut_mixed_2004}: the temperature threshold MLD estimate $h_{MLD}^{\Delta \theta}$ captures a shallower mixed layer as the first thermocline is compensated by salinity effects and does not imprint on the density profile. The resulting pycnocline is indeed deeper than the thermocline is these regions. A few regions of the subtropics, including the Eastern Pacific mode water regions also show an UOP upper limit deeper than 25 m compared to the minimum of the MLD estimates. In these regions, the presence of salinity barrier layers not captured by the UOP method is also possible \citep[e.g.,][]{de_boyer_montegut_control_2007}.

We remind here that the peak detection threshold is based on the vertical standard deviation of the stratification profile. Thus, the detection of peak close to the surface may fail when its change in density is much smaller relative to the change in density of the main pycnocline. Missing such a peak may happen in the Pacific Warm Pool when the change in density due to the salinity barrier layer is much smaller than the main equatorial pycnocline, as illustrated by the stratification profile in the PWP region of Figure~\ref{fig:profile_regions}. 

On the opposite, the detection of very small stratification peaks may happen due to the same definition of the peak threshold: when the profile is very weakly stratified over the whole 0--2000 m depth range, the stratification variance is small and leads to the detection of weak stratification peaks. In such cases, the UOP upper boundary may be located above the MLD estimates as it happens on average in the Labrador sea and in the Nordic seas in winter (Figure \ref{fig:MLD_vs_UOP}b, see also the stratification profile in the region LS of Figure\ref{fig:profile_regions}). This method drawback could be overcome by trying to impose a lower limit to the threshold when the stratification variance is small, but it remains to be tested in the future.

\subsection{UOP and upper ocean stratification measures}

One motivation for this study was the way upper ocean stratification trends have been evaluated in climate studies \citep[e.g.,][]{sallee_summertime_2021, yamaguchi_trend_2019}. Here, we aim to investigate the capacity of two quantities that have been used to capture the relevant upper ocean stratification peak, i.e. the UOP. Using our UOP climatology, we evaluated the percentage of profiles that corresponds to a UOP peak located in the 15-m layer thickness below a MLD defined with a density threshold of 0.03 $kg.m^{-3}$ ($h_{MLD}^{\Delta \sigma}$). We found that this method only captured the UOP peak at best 57.4\% of the profiles in June and at worst 44.9 \% of the profiles in September (Table~\ref{tab:profile_percentage}). 

We also evaluated the number of profiles for which the UOP depth is shallower than 200 m, so that the UOP is included in the density difference between 0 and 200 m. We found that this method captured more regularly the UOP peak for at best 99\% of the profiles in May and at worst 91.9\% of the profiles in September (Table~\ref{tab:profile_percentage}). This method generally fails to capture the UOP peak in winter along the ACC and at high latitudes of the North Atlantic basin (Figure~\ref{fig:uop_characteristics}c).

\begin{table}
\begin{tiny}
  \begin{tabular}{l||c|c|c|c|c|c|c|c|c|c|c|c|}
   & Jan. & Feb. & Mar. & Apr. & May & Jun. & Jul. & Aug. & Sep. & Oct. & Nov. & Dec. \\
  \hline
  $UOP \in [h_{MLD}^{\Delta \rho_\theta}, h_{MLD}^{\Delta \rho_\theta} + 15m]$ & 50.5 & 46.5 & 47. & 51.2 & 56.2 & 57.4 & 53.7 & 47.9 & 44.9 & 47.4 & 52.9 & 54.2 \\
  $UOP \in [0m, 200m]$ & 96.7 & 94.8 & 94.8 & 97.3 & 99. & 97.8 & 94.8 & 91.9 & 91.9 & 95.2 & 98.4 & 98.5
  \end{tabular}
\end{tiny}
  \centering
  \caption{Percentage of profiles when the UOP peak falls into the 15-m layer below the mixed layer (first row), and into the first 200 m (second row).}
  \label{tab:profile_percentage}
\end{table}

Although the density  between 0 and 200 m succeeds in capturing the UOP in most of the oceanic regions, \cite{somavilla_warmer_2017} argued that it does not provide any information about the tendency of the water column to be mixed, linked to the shape of the upper ocean density profile. By definition, the upper ocean stratification index (UOSI) is able to capture the characteristics of the density profiles as it integrates density variations down to the UOP bottom. Note that the UOSI is interpreted as the amount of buoyancy needed to destratify the upper ocean rather the tendency of the water column to be mixed.

\subsection{UOP and the permanent ocean pycnocline}

\begin{figure}[h!]
  \begin{center}
    \includegraphics[width=15cm]{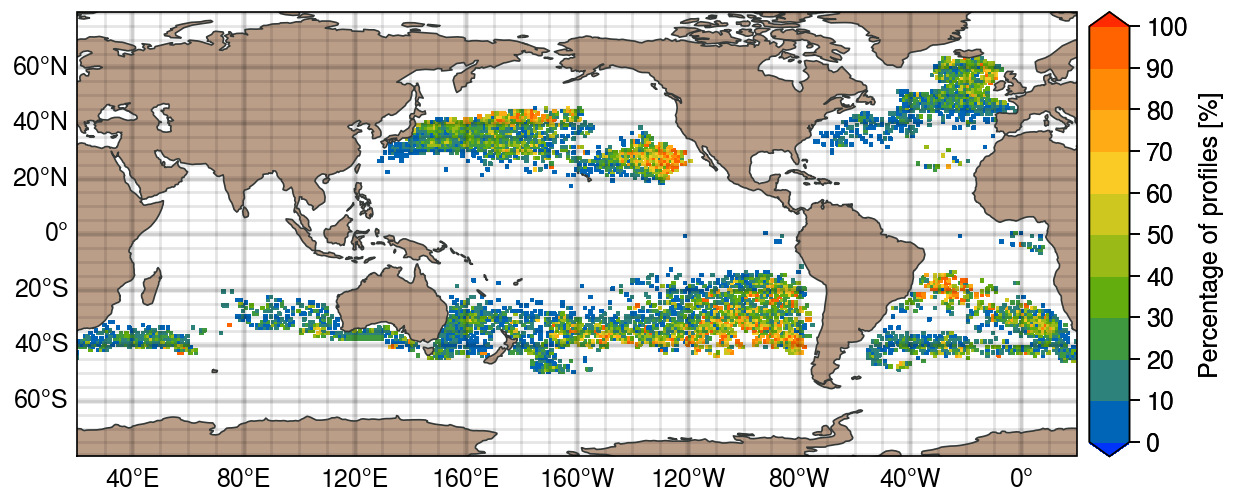}
  \end{center}
  \caption{Percentage of winter profiles whose UOP merges with the permanent pycnocline taken from \citep{feucher_subtropical_2019}.}
\label{fig:UOP_vs_POP}
\end{figure}

Our current UOP climatology does not make any distinction with the permanent ocean pycnocline (POP) so that we are not able to quantify accurately the occurrences when the UOP merges with the POP. However, we compared the POP climatology with the UOP distribution in some regions of the global ocean and we found a clear overlap in winter between the UOP depth and the POP location. To go a bit further, we computed the percentage of winter profiles in each bins when the UOP depth falls into the POP layer as estimated by \cite{feucher_subtropical_2019}. Results are shown in Figure~\ref{fig:UOP_vs_POP}. In the northern hemisphere, merging between the UOP and the POP might happen in the Kuroshio Extension, in the NPESTMW region, and around the Rockall Plateau. The NPESTMW region has the highest percentage of profiles meeting the POP layer. In the southern hemisphere, the SPESTMW region, the Agulhas Leakage and the region offshore of Brazil fed by the south equatorial current show large percentage of profiles with the UOP likely to merge with the POP layer.

This comparison complements the stratification configurations illustrated in section~\ref{sec:descriptive_approach} from the gridded T/S climatology. It also provides additional insights about the relative positions of the upper and the permanent pycnoclines and how deep the ocean surface may communicate with the ocean interior. However, the exact distance between the UOP and the POP remains to be quantified on each profile. Finding the POP could be done based on its smaller variability over time and space and on signatures complementary to the stratification peaks, such as the temperature or the salinity at the bottom of the peak that would reflect the interior ocean. This work is left for the future.

\subsection{Processes setting the UOP thickness}

While the full dynamics of the OSBL is complex and involve several oceanic processes \citep[e.g.,][]{ferrari_eddy-mixed_2004, johnston_observations_2009}, one result of our study is that the UOP thickness is relatively constant over the global ocean. Is there a general physical process that might set this thickness? The existing literature suggests that two main regimes may drive vertical mixing and entrainment in forced stratified shear flow \citep{caulfield_layering_2021, smith_turbulence_2021}, such as those occurring at the base of the mixed layer. A weak stratification may favour a shear-driven overturning regime involving Kelvin-Helmoltz-like instabilities that broadens the stratification interface, while a strong stratification may favour a scouring regime of Holmboe-like instabilities that sharpens the stratification interface. While a laboratory experiment showed that the transition between Kelvin-Helmoltz and Holmboe instabilities existed in the physical world \citep{hogg_kelvinhelmholtz_2003}, recent direct numerical simulations explored the transition between these two kinds of instabilities and highlighted an hybrid regime mixing, characteristics of both instabilities \citep{smith_turbulence_2021}.

\cite{dohan_mixing_2011} noticed two different behaviours of the OSBL following two consecutive storms. The first storm was associated with a deepening of the MLD and a sharpening in the UOP thickness (from 35 m to 25 m), with the stratification effects being more important than the shear effects (i.e., a gradient Richardson number larger than unity). The second storm was associated with little change in the MLD but a broadening of the UOP (from 25 m to 45 m) with a dominance of shear effects over stratification  (i.e., a gradient Richardson number larger than 0.6). The OSBL turbulence regime during the first storm seems dominated by external mixing with potentially scouring effects at the interface, while the turbulence regime during the second storms seems to be internally driven by shear instabilites. Using a Lagrangian float to study instabilities in the region, \cite{kaminski_high-resolution_2021} showed that intermittent Kelvin-Helmoltz instabilities indeed drives intermittent vertical mixing with substantial effects of both stratification and shear leading to a UOP thickness of 10-20 m.

The change in UOP thickness due to different turbulent regimes is potentially small compared to typical vertical scales due to other phenomena occurring in the ocean (e.g., internal waves, submesoscale features). This hampers to draw any conclusions on the dynamical regimes from the unique knowledge of the UOP thickness. Using additional measurements of shear and turbulent dissipation would be a solution, but they are only available in a few spots of the global ocean and a global coverage of these quantities seems out of reach. Some regions where the UOP thickness departs significantly from the global UOP thickness, such as areas around the Rockall Plateau, would however benefit to have these additional measurements. In this region, we found that the stratification was weak in winter ($\mathcal{O}(10^{-5}s^{-2})$, Figure~\ref{fig:2d_distribution}e) and that the UOP thickness was larger than average ($\sim 50$ m, Figure~\ref{fig:uop_characteristics}e). Thus, it would plausible that an overturning regime would be favoured due to a small stratification rather than a scouring regime, which would be consistent with a broadening of the stratification peak.

\section{Conclusion}\label{sec:conclusion}

In this study, we have built a monthly seasonal climatology of the UOP properties based on the analysis of stratification peaks applied on normalised ARGO profiles. Regardless of the season, the UOP is defined as the shallowest significant stratification peak captured by the method, whose detection threshold is proportional to the standard deviation of the stratification profile. The lower and upper boundaries of the UOP are defined at mid height of the peak, and are consistent with geometric features of the density profiles. We summarise here our main findings on the UOP characteristics (depth, intensity and thickness).

We found that the UOP stratification is the largest in the intertropical band ($10^{-3}\,s^{-2}$), and it decreases with latitudes in winter with values down to $10^{-6}\,s^{-2}$ in the ACC and at high latitudes of the North Atlantic. The deepest UOP are also found in similar regions in winter with UOP depth generally deeper than 200 m and reaching up to 2000 m. The UOP tends to shallow towards the equator to meet the depth of the equatorial pycnocline. While the UOP is essentially shallow ($< 50\,m$) and very stratified in summer ($>10^{-4}\,s^{-2}$) in the global ocean, it remains moderately deep ($ 70-80\,m$) and moderately stratified ($\sim 10^{-4}\,s^{-2}$) during this season along the ACC. Largest seasonal amplitudes of the UOP are found in the Kuroshio Extension, along the ACC and at mid-to-high latitudes of the North Pacific. Regional variability is also present near the equator  with a east-west asymmetry reflecting the equatorial main pycnocline. Results from the comparison between the UOP and MLD estimates suggested that $h_{UOP}^{+}$ is a correct proxy for the MLD depth in summer for latitudes higher than $15 ^\circ$, and is also good approximation in winter for the same latitudes except at high latitudes of the North Atlantic.

Looking at specific regions, we highlighted regional variations in the seasonal cycle, that does not occur at the same pace depending on the region, with sometimes abrupt change in UOP depth and sometimes more gradual changes. The UOP variability, including submonthly, interannual and small-scale variability, was dependent on the region, with large variability noticed in the Kuroshio Extension, along the ACC, at mid-to-high latitudes of the North Pacific and in the eastern Pacific mode water regions. Merging between the UOP and the POP seems very likely in the northern hemisphere during winter.

We found that the UOP thickness is relatively constant over the global ocean with a median value of 23 m, except in a few places. In particular, the winter UOP thickness may be much larger than the summer UOP around the Rockall Plateau, but the winter thickness does not exceed 55 m on average. We showed that taking the average of $N^2$ in a 15-m layer below the MLD did not capture the stratification peak everywhere. Thus, the increasing trend of the stratification metric found by \cite{sallee_summertime_2021} may not reflect the full influence of climate change of the upper ocean stratification. What exactly sets the UOP remains an open question but Kelvin-Helmholtz-like and Holmboe-like instabilities involved in stratified shear flows could play a role, depending on the UOP stratification intensity.

We eventually found that the restratification could be captured with the UOSI, a stratification index that we defined based on the UOP properties. This quantity integrates the information on UOP depth, intensity and thickness, and corresponds to the total amount of buoyancy loss required to erode the UOP. The UOSI is generally minimum in spring and associated with a maximum in relative variability during the same season. A focus on a few regions has shown that restratification happens generally in late winter/early spring, but this transition is not homogeneous between regions: as for the depth-intensity seasonal cycle, restratification may occur abruptly or in a smoother way.

The UOP climatology only includes data from ARGO floats for the moment but should be improved by including additional data from other observing systems such those on marine mammals. Moreover, we only provide a monthly climatology of the dataset, but a montlhy-varying dataset could be produced using an optimal interpolation as done for the ISAS gridded dataset. Such a monthly dataset would also permit to disentangle sub-monthly variability from interannual variability, as opposed to the monthly climatology. Combining the UOSI climatology with a surface buoyancy flux climatology would help to estimate the contribution of lateral buoyancy fluxes -- including those due to submesoscales -- to the upper ocean restratification \cite[e.g,][]{johnson_global_2016}, and to analyse in details the large variability of upper ocean stratification occurring in spring. Finally, the sensitivity of our method to defining a lower limit for the detection threshold should be investigated to better understand differences between the upper limit of the UOP and the MLD estimates in the North Atlantic.

\bibliographystyle{unsrtnat}
\bibliography{serazin2022} 

\newpage

\newpage

\newpage

\newpage

\newpage

\newpage

\newpage

\newpage

\newpage

\newpage

\newpage







\end{document}